\documentclass[10pt,superscriptaddress,notitlepage]{revtex4-1}
\usepackage{amsmath,amsfonts}
\usepackage{bbm}
\usepackage{graphicx}
\usepackage{color}
\usepackage{soul}

\begin{document}

\title{Electron-hole coherent states for the Bogoliubov-de Gennes equation}
\author{Sven Gnutzmann}
\email{sven.gnutzmann@nottingham.ac.uk}
\affiliation{School of Mathematical Sciences, University of Nottingham, UK}

\author{Marek Ku{\'s}}
\email{marek.kus@cft.edu.pl} \affiliation{Center for Theoretical Physics,
Polish Academy of Sciences, Warsaw, Poland}

\author{Jordan Langham-Lopez}
\email{pmxjal@nottingham.ac.uk}
\affiliation{School of Mathematical Sciences, University of Nottingham, UK}
\begin{abstract}
  We construct a new set of generalized coherent states, {\em the electron-hole
  coherent states}, for a (quasi-)spin particle on
  the infinite line. The definition is inspired by applications to the Bogoliubov-de
  Gennes equations where the quasi-spin refers to electron- and
  hole-like components of electronic excitations in a superconductor.
  Electron-hole coherent states generally entangle the space
  and the quasi-spin degrees of freedom. We show that the electron-hole
  coherent states allow obtaining a resolution of unity and form minimum
  uncertainty states for position and velocity where the velocity operator is
  defined using the Bogoliubov-de Gennes Hamiltonian.
  The usefulness and the limitations of electron-hole coherent states
  and the phase space
  representations
  built from them are discussed in terms of
  basic applications to the Bogoliubov-de Gennes equation such as
  Andreev reflection.
\end{abstract}


\maketitle

\section{Introduction}

{Coherent} states in their multiple variants and
generalizations have become an indispensable tool in various branches of
quantum mechanics and quantum field theory. The standard coherent states are
strongly connected to the quantum description of a one-dimensional harmonic
oscillator. They were first mentioned by Schr\"odinger \cite{schrodinger26},
who realised that they describe wave packets with minimal uncertainty that
follow the classical dynamics without changing their shape. Glauber
\cite{glauber63,glauber63a} and Sudarshan \cite{sudarshan63} later rediscovered
these states, gave a very detailed description of their properties and showed
their relevance in quantum optics. Since then they have also
become an important tool in quantum field theory and a central ingredient of
phase space formulations of quantum mechanics
and semiclassical approximations.\\
The standard coherent states have been generalised along various lines.
E.g., the group-theoretic approach generalizes the fact that standard
coherent states can be obtained by a group action of the
Heisenberg-Weyl group on the harmonic oscillator ground state.
Generalized coherent states can then be obtained by taking a different
initial state or a different group (and a Hilbert space which carries a
representation of that group) \cite{perelomov12}.

There are various other approaches to generalize coherent states that
are discussed in the literature with many interesting applications
\cite{klauder85,ali14,gazeau09,combescure12}. A property shared by most
generalized coherent states is that they are labeled by continuous
parameters and allow obtaining a formal resolution of unity (\emph{via}
a projector valued
measure on the set of continuous parameters).\\
In this contribution we will consider a special case of
a construction of coherent states on a Hilbert space which is
a tensor product $\mathcal{H}= \mathcal{H}_1 \otimes \mathcal{H}_2$.
There is an obvious canonical construction if each of the factor
spaces already comes with a set of coherent states, say
$ |\alpha \rangle_1 \in \mathcal{H}_1$ and
$ |\beta \rangle_2 \in \mathcal{H}_2$ where $\alpha$ and $\beta$ stands for the
set of continuous parameters on that space. One may then define the coherent
states on the tensor product $\mathcal{H}$ as
the tensor product of coherent states in each factor
$|\alpha,\beta\rangle := |\alpha \rangle_1  \otimes |\beta \rangle_2$.
In the following we will show that the special case where
$\mathcal{H}_1 \equiv L^2(\mathbb{R})$ and
$\mathcal{H}_2 \equiv \mathbb{C}^2$ one may define
generalized coherent states that are (for almost all values of the continuous
parameters) entangled, i.e. not product states.
Physically, the tensor coherent states are quite natural for describing a
quantum point-particle on the line with spin $1/2$ (a point
particle in $\mathbb{R}^n\equiv \mathbb{R}\otimes \mathbb{R} \otimes \dots
\otimes \mathbb{R}$ is described analogously by appropriate
tensor-product coherent states).\\
It is however interesting to see that entangled coherent states,
sharing many properties of the standard coherent states can be defined.
We adopt the name \emph{electron-hole coherent states} for our
construction as they may be used in a natural way for electronic
excitations in a superconductor or a hybrid
superconducting-normalconducting device. These are described by the
Bogoliubov-de Gennes equation and the factor $\mathbb{C}^2$ of the
Hilbert space refers to electron-like and hole-like components of the
wave function.

In Section \ref{can-cs} we review the main properties
of standard coherent states, $SU(2)$ coherent states for arbitrary spin,
and tensor
product states of standard coherent states with the spin
coherent states.\\
In Section \ref{eh-cs} we construct electron-hole coherent states, derive their main properties, and show that they fit
into the group theoretic approach to generalized coherent states if one drops
the assumption that the group has to act linearly on the Hilbert space. We also give a brief account of the
relation of electron-hole coherent states to the Bogoliubov-de Gennes equation
\\
In Section \ref{eh-cs-app} we look at
generalisations of various coherent state representations
of operators and states. We also discuss some limitations of electron-hole
coherent states when trying to generalize the phase space formulations of
quantum mechanics that is based on Husimi functions. These limitations
may be less severe in a semiclassical regime
of the Bogoliubov-de Gennes equation.

\section{A short review of coherent states}
\label{can-cs}

\subsection{Standard (Schr\"odinger-Glauber-Sudarshan) coherent states}

The construction of standard coherent states can be found in various textbooks.
As they are one ingredient of the definition of electron-hole coherent states
in Section \ref{eh-cs} we will give a short
summary of their construction and their basic properties. We will focus mainly
on properties that may be used as alternative definitions -- each of these has
been used as a starting point for generalising the concept of coherent states.
Usually focussing on one property of coherent states will imply that not all
properties of standard coherent states will hold in the generalisation. In
Section \ref{eh-cs} we will analyse how these properties generalise to the
electron-hole coherent states defined there (and what limitations arise).

\subsubsection{The unitary representation of the Heisenberg-Weyl group}

One may start from
the (unique up to isomorphisms) irreducible unitary representation of the
Heisenberg-Weyl group $H_3(\mathbb{R})$. In this representation the group is generated by the annihilation operator
$a$, the creation operator $a^\dagger$ and the identity operator
$\hat{1}$ (if $c$ is a scalar we will usually write $c \hat{1} \equiv c$)
with commutation relations
\begin{equation}
  \left[ a, a^\dagger \right ] = \hat{1} \quad\text{and} \quad
  \left[\hat{1},a\right]=\left[\hat{1},a^\dagger\right]=0 .
\end{equation}
In the unitary representation the creation and annihilation operators are
Hermitian conjugates of each other (in general we will denote
the
Hermitian conjugate of an operator $O$ as $O^\dagger$).\\
The unitary representation of the Heisenberg-Weyl group is obtained by taking
the exponential of  anti-Hermitian linear combinations of the
generators, i.e. a (representation of a) group element is of the form
\begin{equation}
  D(\alpha,\phi)= e^{\alpha a^\dagger -\alpha^* a + i\phi}=
  e^{-\frac{|\alpha|^2}{2}+ i\phi}  e^{\alpha a^\dagger} e^{-\alpha^* a},
\end{equation}
where $\alpha \in \mathbb{C}$ and $\phi \in \mathbb{R}$.
The second equality can be established using the
Baker-Campbell-Hausdorff formula which also allows one to find the
multiplication law
\begin{equation}
  D(\alpha_2,\phi_2) D(\alpha_1,\phi_1)= D(\alpha_1+\alpha_2,
  \phi_1+\phi_2 + \mathrm{Im}(\alpha_1^* \alpha_2)),
  \label{hw-group}
\end{equation}
which is indeed the multiplication law of the Heisenberg-Weyl group $H_3(\mathbb{R})$.
One also finds
\begin{equation}
  D(\alpha,\phi)^{-1}=D(-\alpha,-\phi)= D(\alpha,\phi)^\dagger ,
\end{equation}
which shows that the representation is indeed unitary.\\
The infinite dimensional Hilbert space
$\mathcal{H}_\infty\equiv \ell^2 \equiv L^2(\mathbb{R})$ of this irreducible
representation is spanned by
the orthogonal (and normalised) number state basis (aka Fock basis) $\left\{ |n\rangle \right\}_{n=0}^\infty$
(where orthonormality implies $\langle n | n'\rangle = \delta_{nn'}$)
in which the generators act as
\begin{subequations}
\begin{align}
  a |n\rangle = & \sqrt{n} | n-1\rangle \qquad \text{where $a|0\rangle=0$ for $n=0$,} \\
  a^\dagger |n\rangle =& \sqrt{n+1} |n+1\rangle,\\
  \hat{1} |n\rangle = & |n\rangle, \\
  a^\dagger a |n\rangle =& n |n\rangle.
\end{align}
\end{subequations}
The last equation shows that $|n\rangle$ is an eigenstate of the
number operator $N:= a^\dagger a$. These states are thus the energy eigenstates
the harmonic oscillator which is described by the Hamiltonian $H=\hbar \omega
\left(a^\dagger a + 1/2 \right)$. For $n=0$ the number state
$|0\rangle$ will be called the vacuum.\\
It is well known that the abstract Hilbert space $\mathcal{H}_\infty$
together with the representation of the Heisenberg-Weyl generators
are isomorphic to the Hilbert space of square-integrable functions on
the line $L^2(\mathbb{R})$ which is used to describe quantum point
particle in one dimension. The solution of  the harmonic oscillator provides
a standard way to identify the two Hilbert spaces and representations -- the isomorphism is
however not unique and we will come back to this point.

\subsubsection{The definition of coherent states and some basic properties}

The standard coherent states on $\mathcal{H}_\infty$ can now be
defined by
acting with elements of the Heisenberg-Weyl group on the vacuum
\begin{equation}
  |\alpha\rangle
  := D(\alpha,0) |0\rangle=
  e^{-\frac{|\alpha|^2}{2}} e^{\alpha a^\dagger} |n=0\rangle \ .
  \label{cs_definition}
\end{equation}
Note that $D(\alpha,\phi)|0\rangle = e^{i \phi} |\alpha
\rangle$, so by setting $\phi=0$ in our definition we
fixed a phase convention. It is sometimes convenient to define
coherent states that omit the normalization factor $e^{-|\alpha|^2/2}$,
that is
\begin{equation}
  \left|\alpha\right):=  e^{\alpha a^\dagger} \left|n=0\right\rangle
  = e^{|\alpha|^2/2} \left|\alpha \right\rangle\ .
\end{equation}
While  $\left|\alpha\right)$ is not normalised, it is
analytic as a function of $\alpha$. As a manifold the coherent states are
equivalent to the complex plane $\mathbb{C}$ which is equivalent to the coset
space $H_3(\mathbb{R})/U(1)$ (\emph{via} the map $\alpha \mapsto D(\alpha,0)$).
One may identify this plane with the classical phase space and the
Heisenberg-Weyl group  translates points in phase
space, i.e.
\begin{equation}
  D(\alpha_2,0) |\alpha_1\rangle
  = e^{i \mathrm{Im}(\alpha_1^*\alpha_2)}
  |\alpha_1 +\alpha_2\rangle
  .
\end{equation}
In the rest of the paper standard coherent states in the Hilbert space
$\mathcal{H}_\infty$ will always be denoted with the letter $\alpha$ (sometimes
with an additional index). Note that the vacuum is the coherent state with
$\alpha=0$, i.e. $|\alpha=0\rangle= |n=0\rangle$ (writing just $|0\rangle$ does
thus not lead to ambiguities). A general
coherent state is a superposition of number states
\begin{equation}
  \left|\alpha\right)= \sum_{n=0}^\infty
  \frac{\alpha^n}{\sqrt{n!}} |n\rangle\
  \quad \text{or} \quad
  |\alpha\rangle
  = e^{-\frac{|\alpha|^2}{2}} \sum_{n=0}^\infty
  \frac{\alpha^n}{\sqrt{n!}} |n\rangle\ .
\label{expansion_numberstates}
\end{equation}
This implies that the probability $p_n$ to find the $n$-th number state
in a given coherent state is Poisson distributed $p_n=
\frac{|\alpha|^2}{n!}e^{-|\alpha|^2}$. The probabilities take dominant
values when $n$ is comparable to $|\alpha|^2$. Indeed the maximal
probability arises when $n$ is the largest integer smaller than
$|\alpha|^2$. It is also straightforward to find the expectation value
of the number operator
\begin{equation}
  \langle \alpha|N|\alpha
  \rangle
  =\sum_{n=0}^\infty n p_n = |\alpha|^2 .
\end{equation}
This calculation becomes even simpler if one establishes that standard coherent
states are eigenstates of the annihilation operator,
\begin{equation}
  a |\alpha\rangle
  = \alpha  |\alpha\rangle
  \ .
  \label{cs_eigenequation}
\end{equation}
This property can be used also as an alternative definition for
standard coherent states.
Using the identities
\begin{subequations}
  \begin{align}
    D(-\alpha,0) a D(\alpha,0)=& a +\alpha,\\
     D(-\alpha,0) a^\dagger D(\alpha,0)=& a^\dagger +\alpha^*,
  \end{align}
\end{subequations}
one can derive \eqref{cs_eigenequation}  and show that a definition
based on \eqref{cs_eigenequation}  is equivalent to our definition \eqref{cs_definition} (up to choice
of phase conventions).

Apart from the algebraic structure given so far the probably most
important property of coherent states is that they form a complete set
in $\mathcal{H}_\infty$  that can be used in a similar way as a basis.
Using either the expansion in number states
\eqref{expansion_numberstates} or the group multiplication
\eqref{hw-group} one can establish that coherent states have a
non-vanishing overlap,
\begin{equation}
  \left(\alpha_1|\alpha_2\right)=
  e^{\alpha_1^*\alpha_2}, \qquad
  \langle \alpha_2 |\alpha_1\rangle
  = e^{-\frac{|\alpha_1-\alpha_2|^2}{2} - i
    \mathrm{Im}(\alpha_1^*\alpha_2)} \neq 0,
\end{equation}
which implies that one cannot build an orthonormal basis from coherent
states. Nevertheless, the coherent states are complete (indeed
overcomplete) as they allow obtaining the resolution of unity in terms
of projectors onto coherent states
\begin{equation}
  \hat{1}= \frac{1}{\pi} \int_{\mathbb{C}} d^2\alpha\
    \left|\alpha\right\rangle
    \left\langle \alpha\right|
    = \frac{1}{\pi} \int_{\mathbb{C}} d^2\alpha\ e^{-|\alpha|^2}
    \left|\alpha\right)
    \left( \alpha\right|
    \label{cs_resolution}
\end{equation}
where $d^2\alpha= d\mathrm{Re}(\alpha) d\mathrm{Im}(\alpha)$. This resolution
can be derived either by formally reducing it to the resolution
$\hat{1}=\sum_{n=0}^\infty |n\rangle \langle n|$ in terms of the orthonormal
basis of number states or (more elegantly but also by means of
more advanced tools) using Schur's lemma from representation
theory. From this identity it follows that we can calculate
the trace of an operator $O$ as
\begin{equation}
  \mathrm{tr}\, O = \frac{1}{\pi} \int_{\mathbb{C}} d^2\alpha\
  \langle
  \alpha| O | \alpha \rangle,
\end{equation}
and that we may represent any abstract state $|\psi\rangle \in
\mathcal{H}_\infty$
in terms of an explicit  function
$\psi(\alpha,\alpha^*)\equiv \langle \alpha |\psi\rangle $ on the plane.
The scalar product of two states is then given by
\begin{equation}
  \langle \psi_2| \psi_1 \rangle= \frac{1}{\pi} \int_{\mathbb{C}} d^2\alpha\
  \psi_2(\alpha,\alpha^*)^* \psi_1(\alpha,\alpha^*).
\end{equation}
Note that
$\psi(\alpha,\alpha^*)$ is not an analytic function of $\alpha$. However
$f(\alpha)=\psi(\alpha,\alpha)^* e^{|\alpha|^2/2}\equiv\langle \psi|\alpha)$ 
turns out to be analytic in $\alpha$ and
this gives rise to the so-called Bargmann representation  
(see \cite{perelomov12}).

\subsubsection{Coherent states as minimum uncertainty states}

As $a$ and $a^\dagger$ are not Hermitian operators they do not qualify
as quantum observables. We thus introduce the formal (dimensionless)
position and momentum operators
\begin{subequations}
  \begin{align*}
    Q=& \frac{a+a^\dagger}{\sqrt{2}},\\
    P=&\frac{a-a^\dagger}{\sqrt{2} i},
  \end{align*}
\end{subequations}
which are Hermitian by definition and obey the commutation
relation
\begin{equation}
  i \left[P,Q \right]=\hat{1}\ .
\end{equation}
The operators $Q$, $P$ and $\hat{1}$ may be used as an alternative set
of generators for the Heisenberg-Weyl group $H_3(\mathbb{R})$.
Their expectation values in a coherent state are given  by
\begin{equation}
  \langle \alpha |Q|\alpha\rangle
  =
  \sqrt{2}\, \mathrm{Re}\, \alpha \quad \text{and} \quad
  \langle \alpha |P|\alpha\rangle
  =
  \sqrt{2}\, \mathrm{Im}\, \alpha,
\end{equation}
which shows that we may thing of the alpha plane as the classical phase
space, such that for each point in phase space there is one coherent
state with the corresponding expectation values for momentum and
position.
\\
The uncertainty of an
observable $O$ in the quantum state $|\psi\rangle$ may be measured by
the variance
 $ \mathrm{Var}\left[ O \right]= \langle \psi|O^2|\psi\rangle -\langle
  \psi|O|\psi \rangle^2$
which vanishes if $O$ takes a sharp value in the state $|\psi\rangle$.
Quantum mechanics forbids that $Q$ and $P$ both take sharp values at the
same time in any quantum state $|\psi\rangle$.
Heisenberg's uncertainty principle states
\begin{equation}
  \mathrm{Var}\left[P\right] \mathrm{Var}\left[Q\right] \ge \frac{1}{4}
\end{equation}
for any state $|\psi\rangle$. Standard coherent states obey
\begin{equation}
  \mathrm{Var}\left[P\right] = \mathrm{Var}\left[Q\right]
  =\frac{1}{2}
\end{equation}
and thus minimize the uncertainty. It leads to an alternative
definition of standard coherent states as the
set of minimal uncertainty states with equal variances
of $Q$ and $P$. The minimal uncertainty of coherent
states is one of the main reasons why coherent states have become such an
important tool for investigating quantum-classical correspondence in the
semiclassical regime, when a
coherent state deforms mildly during the dynamics for sufficiently small times.
If one starts the dynamics in the coherent state
$|\alpha_0\rangle$ then for sufficiently short times the state
is well approximated by a trajectory $|\psi(t)\rangle \approx
e^{i\phi(t)}|\alpha(t)\rangle$ of coherent states, where
$\alpha(t)=\frac{q(t)+ip(t)}{\sqrt{2}}$ is a classical trajectory in phase
space (obtained from corresponding classical Hamiltonian dynamics). If the
classical dynamics is not chaotic this approximation may work for quite long
times. A special case is the harmonic oscillator $H=\frac{\omega}{2}\left(P^2 +
Q^2 \right)=\omega \left(a^\dagger a + \frac{1}{2}\right)$, where $e^{-i Ht}
|\alpha_0\rangle = e^{-i\omega t/2} |\alpha(t)\rangle$ remains exact for the
classical trajectory $\alpha(t)= e^{-i \omega t} \alpha_0$. In the position
representation this leads to a Gaussian function whose centre follows the
classical oscillations in space without changing its shape otherwise. This is
the form in which coherent states were used by
Schr\"odinger to investigate quantum-classical correspondence for the harmonic
oscillator \cite{schrodinger26}.

\subsection{$SU(2)$ or spin coherent states}

One well-known generalisation of the standard coherent states replaces
the Heisenberg-Weyl group by some other (physically relevant) group $G$
whose generators are observables of a quantum system. This has been
developed mainly by Gilmore \cite{gilmore72} and Perelomov
\cite{perelomov72, perelomov12}. In this case one takes an irreducible
representation of $G$ on an appropriate Hilbert space, chooses an
appropriate reference state $|\mu\rangle$ and then defines the coherent
states by $|g\rangle = g |\mu\rangle$ where $g\in G$ is an element of
the group. If the reference state is chosen appropriately one may find
coherent states which have analogues of almost all main properties of
the standard coherent states (overcompleteness with an explicit
resolution of unity, appropriately defined minimal uncertainties,
relation to classical phase-space, etc.).

The most prominent example of this are the $SU(2)$ coherent states
\cite{radcliffe71,arecchi72}, which are also known as spin coherent states or
angular momentum coherent states. The generators of
the $SU(2)$ group are the angular momentum or spin operators $J_1$, $J_2$ and
$J_3$ with commutation relations
\begin{equation}
  \left[ J_1,J_2\right]=i J_3 ,\quad \left[ J_2, J_3\right]=i J_1 \quad
  \text{and}  \left[ J_3, J_1\right]=i J_2 .
\end{equation}
The irreducible representations are characterised by the half-integer number
$j$, such that the total angular momentum is $J_1^2+J_2^2+J_3^2=
j\left(j+1/2\right)$ and the dimension of the Hilbert space is $2j+1$. In the
present context only the simplest case $j=1/2$ is relevant, so we will focus
our summary on these (see \cite{perelomov12} for details). In that case the
Hilbert space $\mathcal{H}_2\equiv \mathbb{C}^2$ is spanned by two orthogonal
states,
\begin{equation}
  |+\rangle \equiv \begin{pmatrix} 1\\ 0 \end{pmatrix}
  \quad \text{and} \quad |-\rangle \equiv  \begin{pmatrix} 0\\ 1 \end{pmatrix},
\end{equation}
and the $SU(2)$ generators are represented by the (Hermitian)
Pauli matrices,
\begin{equation}
  J_1\equiv
  \frac{1}{2}
  \sigma_1 =
  \begin{pmatrix}
    0&1\\
    1&0
  \end{pmatrix}, \quad
  J_2\equiv \frac{1}{2}
  \sigma_2 =
  \begin{pmatrix}
    0&-i\\
    i&0
  \end{pmatrix},
  \quad \text{and} \quad
  J_3\equiv \frac{1}{2} \sigma_3 =
  \begin{pmatrix}
    1&0\\
    0&-1
  \end{pmatrix} \ .
  \label{eq:pauli}
\end{equation}
Let us introduce the combinations
\begin{equation}
  J_+=J_1+ i J_2 \equiv  \sigma_+=
  \begin{pmatrix}
    0&1\\
    0&0
  \end{pmatrix}
  \qquad
  J_-=J_1- i J_2 \equiv  \sigma_-=
  \begin{pmatrix}
    0&0\\
    1&0
  \end{pmatrix}\ .
  \label{eq:raising}
\end{equation}
Note that \eqref{eq:pauli} contain a factor $1/2$ that is omitted in
\eqref{eq:raising} (for convenience).
Almost all elements of $SU(2)$ are covered by the
parameterisation
\begin{equation}
  U(\beta,\varphi):=
  \begin{pmatrix}
    \frac{e^{i\varphi}}{\sqrt{1+|\beta|^2}} &
    \frac{-\beta^*e^{-i\varphi}}{\sqrt{1+|\beta|^2}}\\
     \frac{\beta e^{i\varphi}}{\sqrt{1+|\beta|^2}} &
     \frac{e^{-i\varphi}}{\sqrt{1+|\beta|^2}}
  \end{pmatrix}
  \equiv e^{\beta J_-} e^{-\log(1+|\beta|^2) J_3} e^{-\beta^* J_+} e^{i 2\varphi J_3}
\end{equation}
where $\beta \in \mathbb{C}$ and $0 \le \phi \le 2\pi$. The $SU(2)$ coherent states are
now defined by
\begin{equation}
  |\beta\rangle= U(\beta,0)|+\rangle
  = \frac{1}{\sqrt{1+|\beta|^2}} \left(|+\rangle + \beta |-\rangle\right)
    \equiv \begin{pmatrix}
      \frac{1}{\sqrt{1+|\beta|^2}}\\
      \frac{\beta}{\sqrt{1+|\beta|^2}}
    \end{pmatrix}
\end{equation}
to which one should add the coherent state $|\beta=\infty\rangle \equiv
|-\rangle$. Again it is sometimes useful to define spin
coherent states that are analytic in $\beta$ but not normalized by
\begin{equation}
  \left|\beta\right)= e^{\beta J_-}|+\rangle= |+\rangle + \beta |-\rangle\ .
\end{equation}
As a manifold the coherent states  form the sphere
$S^2= SU(2)/U(1)$ (see below for more details), which is the appropriate
classical phase space for the dynamics of an angular momentum vector (with
fixed length).

In analogy to the standard coherent states the $SU(2)$ coherent states can be
used as an overcomplete basis in the Hilbert space
$\mathcal{H}_{2j+1}=\mathbb{C}^{2j+1}$ of the irreducible $SU(2)$
representation with spin $j$. For $j=1/2$ it is straightforward to evaluate the
overlap
\begin{equation}
  \left(\beta_2|\beta_1\right)=1+\beta_2^*\beta_1 \quad \text{or}
  \quad
  \langle \beta_2| \beta_1 \rangle=
  \frac{1+\beta_2^*\beta_1}{\sqrt{\left(1+|\beta_2|^2\right)
      \left(1+|\beta_1|^2 \right)}}
\end{equation}
of two coherent states
and find an explicit resolution of unity as
\begin{equation}
  \hat{1}=\frac{2}{\pi} \int_{\mathbb{C}} d^2 \beta
  \frac{1}{\left(1+|\beta|^2 \right)^2}
  |\beta\rangle \langle \beta | =
  \frac{2}{\pi} \int_{\mathbb{C}} d^2 \beta
  \frac{1}{\left(1+|\beta|^2 \right)^3}
  \left|\beta\right)\left( \beta \right|
\end{equation}
where the integration over $\mathbb{C}$ may be considered as
an intregral over the classical phase space which is a sphere.
This allows us to represent states in $\mathcal{H}_2$ as
complex functions on phase space in analogy to the standard coherent states
described above.\\
One may see that the coherent states span a manifold equivalent to a sphere
by considering the expectation values
\begin{align}
  \langle \beta| \sigma_+|\beta\rangle=& \frac{\beta}{1+|\beta|^2}\ ,&
  \langle \beta| \sigma_-|\beta\rangle=& \frac{\beta^*}{1+|\beta|^2}\ ,\nonumber \\
  \langle \beta| \sigma_1|\beta\rangle=& \frac{2 \mathrm{Re}(\beta)}{1+|\beta|^2}\ ,&
  \langle \beta| \sigma_2|\beta\rangle=& \frac{2 \mathrm{Im}(\beta)}{1+|\beta|^2}\ ,&
  \langle \beta| \sigma_3|\beta\rangle= \frac{1-|\beta|^2}{1+|\beta|^2}\ .
  \label{sigma_expectation}
\end{align}
The three expectation values $\langle \beta|\sigma_j|\beta\rangle$ ($j=1,2,3$)
build a vector in $\mathbb{R}^3$ of unit length,
i.e. $\sum_{j=1}^3 \langle \beta| \sigma_j\rangle^2=1$ and it is clear that
mapping coherent states to points on the unit sphere with
\eqref{sigma_expectation} is one-to-one (including the coherent state
$|\beta=\infty\rangle$ which maps to the south pole).\\
To conclude this section let us comment on the minimum uncertainty
properties of $SU(2)$ coherent states. For the $j=1/2$ representation
every normalized state is a coherent state (up to an irrelevant phase),
so coherent states are not distinguished by any minimal uncertainty.
For other irreducible representations $j\ge 1$ $SU(2)$ coherent states
do fulfill a minimal uncertainty relation (see
\cite{delbourgo77,delbourgo77a,perelomov12}).

\subsection{Product coherent states}
\label{sec:product_cs}

It is straightforward to define coherent
states on the tensor product space $\mathcal{H}_{\otimes}
=\mathcal{H}_\infty \otimes \mathcal{H}_2$ by taking tensor product of coherent
states in the two factors. If $|\alpha\rangle \in \mathcal{H}_\infty$ is a
standard coherent state and $|\beta\rangle \in \mathcal{H}_2$ is a spin
coherent state then one defines a (tensor-)product coherent
states as
\begin{equation}
  |\alpha \otimes \beta \rangle:=
  |\alpha\rangle \otimes |\beta\rangle=
  \left[D(\alpha,0)\otimes U(\beta,0)\right]\, |0\rangle \otimes|+\rangle\ .
  \label{product_cs_definition}
\end{equation}
The analytic but non-normalised variant is
defined as $\left|\alpha \otimes \beta \right):= e^{\alpha a^\dagger}e^{\beta
J_-} |0\rangle \otimes |+\rangle  = \left|\alpha \right) \otimes \left|\beta
\right)$. The product coherent states are associated
with the group $H_3(\mathbb{R}) \times SU(2)$, which acts naturally on
$\mathcal{H}_{\otimes}$, as each factor is equipped with an
appropriate irreducible representation. This means that acting
with any element of $H_3(\mathbb{R}) \times SU(2)$ on a product coherent state
gives a product coherent state up to
an additional scalar phase factor.

Product coherent states of this or analogous forms are often
very useful in the analysis of quantum systems that
involve tensor product spaces. For instance, the states $|\alpha \otimes \beta
\rangle$ are often a natural choice to analyse equations of the form
\begin{equation}
  i \hbar \frac{d}{dt} |\Psi(t)\rangle
  = \left( H_0(P,Q)+ \prod_{j=x,y,z} H_j(P,Q) \sigma_j
    \right) |\Psi(t)\rangle,
\end{equation}
where $H_0(P,Q)$ and $H_j(P,Q)$ are Hermitian operators on $\mathcal{H}_\infty$
lifted naturally to $\mathcal{H}_\otimes$ (such standard abuse of notation will
be used frequently). One important example in this class is the Pauli equation
where $H_0(P,Q)= \frac{1}{2m}(P-e A(Q))^2 + V(Q)$ with the magnetic potential
$A(Q)$, and $H_j(Q,P)= \hbar \mu B_j(Q)$ is the magnetic field that couples to
the spin variables $\sigma_j$. In this case in the leading order of the
semiclassical asymptotics (formally $\hbar \to 0$) the spin degrees of freedom
decouple from the space degrees of freedom and an
initial coherent state $|\alpha_0\otimes \beta_0\rangle$ will approximately
remain a coherent state $|\alpha(t) \otimes \beta(t)\rangle$ such that
$\alpha(t)$ follows the corresponding classical dynamics generated by $H_0$ and
the spin rotates along the trajectory.

It is interesting to note that \eqref{product_cs_definition} is not the only
way to define a continuous overcomplete basis on $\mathcal{H}_\otimes$ that one
may build from the standard coherent states in the factor $\mathcal{H}_\infty$,
as we will show in the next section.

\section{Electron-hole coherent states and their properties}
\label{eh-cs}

Let us define
\emph{electron-hole coherent states} in $\mathcal{H}_\otimes=\mathcal{H}_\infty \otimes \mathcal{H}_2$  as
\begin{equation}
  |\alpha \Join \beta\rangle
  =  \frac{1}{\sqrt{1+|\beta|^2}} |\alpha\rangle \otimes |+\rangle
  +
  \frac{\beta^*}{\sqrt{1+|\beta|^2}} |\alpha^* \rangle \otimes |-\rangle
\
\end{equation}
where $|\alpha\rangle$ is a standard coherent state in the factor $\mathcal{H}_\infty$.
The corresponding non-normalized variant
\begin{equation}
  \left|\alpha \Join \beta\right)
  =  \left|\alpha\right) \otimes |+\rangle
  +
  \beta^* \left|\alpha^* \right) \otimes |-\rangle
\
\end{equation}
will also be used as many formulas turn out to be more compact than by using
the normalised variants.\\
For the normalised states we formally add the point $\beta=\infty$ by defining
$|\alpha \Join \infty\rangle= |\alpha^* \rangle \otimes |-\rangle$.

Note that the states are generally entangled (unless $\beta=0$ or
$\beta=\infty$). They do not seem to arise in an obvious way from a
linear group action. We will show, however, that they possess many
properties associated to coherent states. E.g. they form an
overcomplete basis that allows obtaining a straightforward resolution
of unity with associated coherent state representations of the Hilbert
space and its operators. We will also show that it is indeed associated
with a group action -- though not a linear one, as will be explained in
Section \ref{sec:group_action}.

The name electron-hole coherent states refers to possible applications in
superconducting systems described by the Bogoliubov-de Gennes equation -- in
that setting the basis $\{|+\rangle, |-\rangle\}$ of the factor $\mathcal{H}_2$
refers to electron-like and hole-like excitations in a
superconductor.

\subsection{Properties of electron-hole coherent states}

It is straightforward to evaluate the overlap between two different
electron-hole coherent states or the overlap of a electron-hole coherent state
with a product coherent state
\begin{equation}
  \begin{split}
    \left(\alpha_1 \Join \beta_1 | \alpha_2 \Join \beta_2 \right)=&
    e^{\alpha_1^* \alpha_2} + \beta_1 \beta_2^* e^{\alpha_1 \alpha_2^*}\\
    \left(\alpha_1 \otimes \beta_1 | \alpha_2\Join \beta_2 \right)=&
    e^{\alpha_1^* \alpha_2} + \beta_1^* \beta_2^* e^{\alpha_1^* \alpha_2^*}
  \end{split}
\end{equation}
for the non-normalised states. The corresponding overlaps for the normalised
coherent states follow straightforwardly.

The electron-hole states allow obtaining the following resolutions of
unity
\begin{equation}
  \begin{split}
    \hat{1}=& \frac{2}{\pi^2}
    \int d^2\alpha\ d^2\beta\ \frac{1}{(1+|\beta|^2)^2} \left|\alpha\Join \beta\right\rangle
    \left\langle \alpha\Join \beta \right|\\
    =& \frac{2}{\pi^2} \int d^2\alpha\ d^2\beta\
    \frac{e^{-|\alpha|^2}}{(1+|\beta|^2)^3} \left|\alpha\Join \beta\right)
    \left( \alpha\Join \beta \right|,
  \end{split}
  \label{resolution_unity_eh}
\end{equation}
which can be shown straightforwardly by first integrating out $\beta$ and then
applying the resolution of unity of standard coherent states. The resolution of
unity allows us to represent states and operators in the overcomplete basis of
electron-hole coherent states. We will explore such representations in sections
\eqref{sec:eh-cs-representation} and \eqref{sec:phasespace-rep}.

\subsubsection{Expectation values in electron-hole coherent states}

As we have reviewed above, standard coherent states can be viewed as a mapping
from classical phase space to minimal uncertainty states in a
quantum Hilbert space, such that expectation values of
position and momentum  give back the original phase space point. In this
section we want to discuss the expectation values
of the central quantities and see to which extent
analogous properties hold for electron-hole coherent states. It is
straightforward to evaluate the expectation values of the fundamental
observables. Writing $\alpha=\frac{q+iv}{\sqrt{2}}$ where $q$ and $v$ are real
one obtains
\begin{subequations}
\begin{align}
  \langle \alpha \Join \beta| a | \alpha \Join \beta \rangle =&
  \frac{q}{\sqrt{2}} + i \frac{1-|\beta|^2}{1+|\beta|^2}\frac{v}{\sqrt{2}}
  \\
  \langle \alpha \Join \beta| a^\dagger| \alpha \Join \beta \rangle =&
  \frac{q}{\sqrt{2}} - i \frac{1-|\beta|^2}{1+|\beta|^2}\frac{v}{\sqrt{2}}
  \\
  \langle \alpha \Join \beta| Q |\alpha \Join \beta \rangle =&
  q
  \\
  \langle \alpha \Join \beta| P |\alpha \Join \beta \rangle =&
  \frac{1-|\beta|^2}{1+|\beta|^2} v
  \\
  \langle \alpha \Join \beta| \sigma_+ | \alpha \Join \beta \rangle =&
  \frac{\beta^* e^{-v^2-iqv}}{1+|\beta|^2}
  \\
  \langle \alpha \Join \beta| \sigma_- | \alpha \Join \beta \rangle =&
  \frac{\beta e^{-v^2+iqv}}{1+|\beta|^2}
  \\
  \langle \alpha \Join \beta| \sigma_1 | \alpha \Join \beta \rangle =&
  \frac{ 2e^{-v^2}  \mathrm{Re}\left(\beta e^{iqv}\right)}{1+|\beta|^2}
  \\
  \langle \alpha \Join \beta| \sigma_2 | \alpha \Join \beta \rangle =&
  - \frac{ 2e^{-v^2}  \mathrm{Im}\left(\beta e^{iqv}\right)}{1+|\beta|^2}
  \\
  \langle \alpha \Join \beta| \sigma_3 | \alpha \Join \beta \rangle =&
  \frac{1-|\beta|^2}{1+|\beta|^2}\ .
\end{align}
\end{subequations}
The expectation values of $P$, $\sigma_1$ and $\sigma_2$
do not suggests a clear
connection to a phase space description. Replacing the momentum operator $P$
by the operator
\begin{equation}
  V= P \sigma_3
\end{equation}
makes the picture somewhat nicer as
\begin{equation}
  \langle \alpha \Join \beta| V | \alpha \Join \beta \rangle = v\ .
\end{equation}
We will see later in Section \ref{sec:Bdg}, that $V$ has the interpretation of
a velocity operator in the context of the Bogoliubov-de Gennes equation. Note
that $P=V\sigma_3$ and $P^2 = V^2$. We thus regard $\alpha=
\frac{q+iv}{\sqrt{2}}$ as a point in  the phase space spanned
by position and velocity, rather than position and momentum. The expectation
values of the Pauli-matrices $\sigma_1$ and $\sigma_2$ imply
\begin{equation}
  R^2:=\sum_{j=1}^3 \langle \alpha \Join \beta| \sigma_j| \alpha \Join \beta\rangle^2
  = 1 - (1- e^{-2 v^2}) \frac{4|\beta|^2}{(1+|\beta|^2)^2} \le 1
  \label{R2_ehcs}
\end{equation}
which, in general, does not describe a unit sphere. If we allow the sphere to
deform into an ellipsoid we find indeed
\begin{equation}
  e^{2v^2}\left( \langle \alpha \Join \beta| \sigma_1| \alpha \Join \beta\rangle^2
    +\langle \alpha \Join \beta| \sigma_2| \alpha \Join \beta\rangle^2\right) +
  \langle \alpha \Join \beta| \sigma_3| \alpha \Join \beta\rangle^2
  = 1\ .
\end{equation}
Altogether this implies that the phase space underlying electron-hole coherent
states can be viewed as the plane $\mathbb{R}^2\equiv \mathbb{C}$ spanned by
position and velocity expectation values, and an ellipsoid of revolution with two axes
of length $e^{-v^2}$ and the third one of the unit length, attached at
each point of the plane.

\subsubsection{Minimum uncertainty}

Let us now show that electron-hole coherent states are minimal uncertainty
states with respect to position operator $Q$ and velocity operator $V$. It is
straightforward to show that coherent states have uncertainties
\begin{equation}
  \mathrm{Var}_{|\alpha \Join \beta\rangle}[V]=\mathrm{Var}_{|\alpha \Join \beta\rangle}[Q]=\frac{1}{2}
\end{equation}
and thus
\begin{equation}
   \mathrm{Var}_{|\alpha \Join \beta\rangle}[V]\mathrm{Var}_{|\alpha \Join \beta\rangle}[Q]=\frac{1}{4}\ .
\end{equation}
Now the standard uncertainty relation for two operators $A$ and $B$
and an arbitrary state $|\Psi\rangle$  in quantum
mechanics reads $\mathrm{Var}_{|\Psi\rangle}[A]\mathrm{Var}_{|\Psi\rangle}[B]\ge
\left|
  \langle \Psi| [A,B]|\Psi \rangle
\right|^2/4$. In our case $[V,Q]= -i \sigma_3$ what
results in a lower bound $|\langle \Psi| \sigma_3|\Psi \rangle|^2/4$
taking any value between zero and a
quarter. The stricter uncertainty relation
\begin{equation}
  \mathrm{Var}_{|\Psi\rangle}[V]\mathrm{Var}_{|\Psi\rangle}[Q]\ge \frac{1}{4}
  \label{VQ-uncertainty}
\end{equation}
can be obtained by a slight modification of the
standard derivation by first writing (using $V=\sigma_3P$ and
$\sigma_3^2=\hat{1}$ and defining $v=\langle \Psi| V |\Psi\ \rangle$ and
$q=\langle \Psi| Q |\Psi\ \rangle$ )
\begin{equation}
  \mathrm{Var}_{|\Psi\rangle}[V]\mathrm{Var}_{|\Psi\rangle}[Q]=
  \langle \Psi| \left(V-v\right)^2|\Psi \rangle
  \langle \Psi|\left(Q-q\right)^2|\Psi\rangle
  =
  \langle \Psi| \left(P-v\sigma_3\right)^2|\Psi \rangle
  \langle \Psi|\left(Q-q\right)^2|\Psi\rangle\ .
\end{equation}
We may now introduce operators $\Delta P= P -\sigma_3
v$ and $\Delta Q=Q-q$ and proceed as in the standard derivation, i.e. using
Hermiticity and Schwartz's inequality we write
\begin{equation}
  \langle \Psi|\Delta P^2|\Psi \rangle
  \langle \Psi|\Delta Q^2|\Psi\rangle\ =
  \langle \Delta P \Psi|\Delta P \Psi\rangle
  \langle \Delta Q \Psi|\Delta Q \Psi\rangle
  \ge \left| \langle \Psi| \Delta P \Delta Q |\Psi\rangle \right|^2\ .
\end{equation}
Now, as for any complex number $z$ one has
$|z|^2\ge \left(\mathrm{Im}\ z\right)^2$  and
\begin{equation}
  \mathrm{Im} \langle \Psi| \Delta P \Delta Q |\Psi\rangle
  = \frac{1}{2i} \langle\Psi| [\Delta P, \Delta Q]| \Psi \rangle=\frac{1}{2},
\end{equation}
and the uncertainty \eqref{VQ-uncertainty} follows.

While we have found that electron-hole coherent states obey an uncertainty
relation in $V$ and $Q$, they do not obey any uncertainty relation for the
quasi-spin variables $\sigma_j$. With $\Delta \sigma_j =
\sigma_j - \langle\Psi|\sigma_j| \Psi\rangle$, the uncertainty
relation obeyed by spin coherent states is $\sum_{j=1}^3 \sum \langle
\Psi|\Delta \sigma_j^2|\Psi\rangle\ge 2$. This lower bound for the sum over
uncertainties is equivalent to the upper bound $\sum_{j=1}^3 \langle
\Psi|\sigma_j|\Psi \rangle^2 \le 1$ and we have seen above that the quasi-spin
expectation values do not lie on the unit sphere but on an ellipsoid inside the
sphere. The failure of electron-hole coherent states to obey a minimum
uncertainty relation is related to entanglement between the two factor spaces
$\mathcal{H}_\infty$ and $\mathcal{H}_2$ as we will discuss next.

\subsubsection{Entanglement of electron-hole coherent states}
\label{sec:entanglement}

Let us first consider how quasi-spin expectation values can measure the amount
of mixing for quasi-spin states in $\mathcal{H}_2$. For this note that all pure
(normalised) states $|\xi\rangle\in\mathcal{H}_2$ obey $\sum_{j=1}^3 \langle
\xi| \sigma_j | \xi \rangle^2=1$. A mixed states in $\mathcal{H}_2$ is
described by a density matrix
\begin{equation}
  \rho= p_1 |\xi_1\rangle \langle \xi_1 | + p_2 |\xi_2\rangle \langle \xi_2|
\end{equation}
where $p_1,p_2\ge 0$ and $p_1+p_2=1$. The expectation values of a quasi-spin
matrix $\sigma_j$ are $\mathrm{tr}\, \rho \sigma_j= \sum_{m=1}^2 p_m \langle
\xi_m|\sigma_j|\xi_m \rangle$ Without loss of generality we may assume that
$|\xi_1\rangle$ and $|\xi_2\rangle$ are orthogonal (as $\rho$ is
Hermitian). The state is pure if $\rho^2=\rho$, that is either
$p_1=0$ or $p_1=1$. We want to show that the quantity
\begin{equation}
  R^2:=\sum_{j=1}^3 \left(\mathrm{tr}\, \rho \sigma_j\right)^2
  \label{def_R2}
\end{equation}
equals unity if and only if the state is pure, and
$R^2<1$ otherwise. Since $R^2$ is a rotation invariant quantity we may assume
$|\xi_1\rangle = |+\rangle$ and $\xi_2\rangle= |-\rangle$ and thus obtain
\begin{equation}
  R^2= (p_1-p_2)^2=(1-2p_1)^2\ .
\end{equation}
So $R^2=1$ if and only if $p_1=0$ or $p_1=1$ which is what we
wanted to show. Moreover $R^2<1$ for $0<p_1<1$ with a minimum
at $p_1=1/2$ where $R^2=0$.
We can thus use $R^2$ to measure the amount of mixing in a state $\rho$.\\
Now let us come back to the product space $\mathcal{H}_\otimes= \mathcal{H}_\infty \otimes
\mathcal{H}_2$. A pure state $|\Psi\rangle \in \mathcal{H}_\otimes$ is called
separable if
$|\Psi\rangle= |\phi\rangle \otimes |\xi\rangle$ with $|\phi\rangle \in \mathcal{H}_\infty$
and $|\xi\rangle \in \mathcal{H}_2$. A state $|\Psi\rangle \in \mathcal{H}_\otimes$ is called entangled
if it is not separable.
The product coherent states $|\alpha \otimes \beta\rangle$ discussed in section \ref{sec:product_cs} are
seperable by definition. The electron-hole coherent states $|\alpha \Join \beta\rangle$ are separable
if either  $\beta=0$ (and $\alpha$ arbitrary) or $\beta=\infty$ (and $\alpha$ arbitrary)
or $\mathrm{Im}(\alpha) =v/\sqrt{2}=0$ (and $\beta$ arbitray) -- otherwise they are entangled.\\
We may measure the amount of entanglement of a pure state $|\Psi\rangle\in \mathcal{H}_\otimes$
by considering the reduced density matrix
\begin{equation}
  \rho_\mathrm{red} = \mathrm{tr}_{\mathcal{H}_\infty} |\Psi\rangle \langle \Psi |
\end{equation}
where the trace is taken over the factor $\mathcal{H}_\infty$, so that
$\rho_\mathrm{red}$ is an operator in $\mathcal{H}_2$. It is easy to
see that $|\Psi\rangle$ is separable if and only if
$\rho_{\mathrm{red}}=\rho_{\mathrm{red}}^2$ describes a pure state. The
quantity $R^2 \ge 0$ defined in \eqref{def_R2} applied to
$\rho_{\mathrm{red}}$ can thus be used as a measure of entanglement
such that $R^2=1$ for separable states and $R^2<1$ for entangled states
with maximal entanglement for $R^2=0$. A short calculations shows
\begin{equation}
  R^2= \sum_{j=1}^3 \langle \Psi | \sigma_j | \Psi \rangle^2,
\end{equation}
which we have given for electron-hole coherent states in equation \eqref{R2_ehcs}.
For a fixed value $\alpha=(q+iv)/\sqrt{2}$ electron-hole coherent states
obey $1\ge R^2 \ge  e^{-2v^2}>0$ where the lower bound (maximal entanglement)
is obtained when $|\beta|=1$ (the upper bound is obtained for $\beta=0$ and $\beta=\infty$).

\subsection{The group theoretic approach to electron-hole coherent states}
\label{sec:group_action}

Though there is no obvious group which transforms one electron-hole
coherent state into another \emph{via a linear representation} in
Hilbert space we will demonstrate here that there is a
\emph{non-linear} action of the product group $H_3(\mathbb{R}) \times
SU(2)$. We have seen in Section \eqref{sec:product_cs} that the product
coherent states are spanned by the action of the linear
(tensor-product) representation of this group acting on the state
$|0\rangle \otimes |+\rangle \in \mathcal{H}_\infty \otimes
\mathcal{H}_2$. Repeated linear actions of the group $H_3(\mathbb{R})
\times SU(2)$ transform then product coherent states to product
coherent states (up to a phase factor). A small twist of this action
will render the group action non-linear and lead to an analogous
behaviour for electron-hole coherent states.

Let us first define an anti-unitary time-reversal operator
$\mathcal{T}$ by its action on an arbitrary state in
$\mathcal{H}_\infty \otimes \mathcal{H}_2$
\begin{equation}
  \mathcal{T}\sum_{n=0}^\infty
  \left(a_{n,+}|n\rangle
    \otimes |+\rangle  + a_{n,-} |n\rangle \otimes|-\rangle
  \right)
  =
  \sum_{n=0}^\infty \left( a_{n,+}^*|n\rangle
  \otimes |+\rangle  + a_{n,-}^* |n\rangle \otimes|-\rangle
  \right),
\end{equation}
where $|n\rangle$ is the number basis in $\mathcal{H}_\infty$. Anti-linearity
and anti-unitarity follow straightforwardly from the
definition. Acting with $\mathcal{T}$ on a product coherent state
\begin{equation}
  \mathcal{T} |\alpha \otimes \beta\rangle = |\alpha^* \otimes \beta^*\rangle,
\end{equation}
gives just the product coherent state with
complex conjugate parameters $\alpha$ and $\beta$ ($\beta=\infty$ should be
considered as a real number
for this purpose).\\
Now let us define the operator
\begin{equation}
  \mathcal{Z}= \hat{1}_\infty \otimes |+\rangle \langle+|
  + \mathcal{T} \left(\hat{1}_\infty  \otimes |+\rangle \langle+|\right)
\end{equation}
in $\mathcal{H}_\infty \otimes \mathcal{H}_2$, where $\hat{1}_\infty$
is the identity operator in the first factor. This is neither a linear
nor an anti-linear operator as it behaves linearly in one subspace but
anti-linearly in the other. It is also clear from the definition that
$\mathcal{Z}$ is its own inverse,
\begin{equation}
  \mathcal{Z}^2=\hat{1} \qquad\Rightarrow \qquad  \mathcal{Z}^{-1}=\mathcal{Z}\ .
\end{equation}
Moreover, it leaves the norm of any state $|\psi\rangle \in
\mathcal{H}_\infty\otimes \mathcal{H}_2$ invariant,
\begin{equation}
  \langle \mathcal{Z} \psi | \mathcal{Z}\psi \rangle=
  \langle \psi|\psi\rangle.
\end{equation}
However, in general one has $\langle \mathcal{Z} \psi_1| \mathcal{Z}
\psi_2 \rangle \neq \langle \psi_1|\psi_2\rangle$, so $\mathcal{Z}$
cannot be considered as a unitary operation. Moreover there is no way
to define a generalised adjoint to $\mathcal{Z}$, such an operation
would need to satisfy at least $|\langle \psi_1| \mathcal{Z}
\psi_2\rangle|= |\langle \mathcal{Z}^\dagger \psi_1| \psi_2\rangle|$
for two arbitrary states,
and this leads to contradictions.\\
Our main interest here stems from the fact that $\mathcal{Z}$
transforms product coherent states into electron-hole coherent states
and \textit{vice versa},
\begin{equation}
  \mathcal{Z} |\alpha \otimes \beta\rangle
  = |\alpha \Join \beta \rangle
  \quad \Leftrightarrow
  \quad
   \mathcal{Z} |\alpha \Join \beta\rangle
  = |\alpha \otimes \beta \rangle\ ,
\end{equation}
which implies
\begin{equation}
  |\alpha \Join \beta\rangle= \mathcal{Z}
  \left[D(\alpha,0) \otimes U (\beta)\right] \mathcal{Z}
  |0\rangle \otimes |+\rangle.
\end{equation}
We thus define the action of the group $H_3(\mathbb{R})\times SU(2)$ in
$\mathcal{H}_\infty \otimes \mathcal{H}_2$ by the operator $\mathcal{Z}
\left[D(\alpha,\phi)\otimes U(\beta,\varphi)\right]\mathcal{Z}$, where
$D(\alpha,\phi)$ and $U(\beta,\varphi)$ are the corresponding unitary
representations of the two groups in the corresponding spaces. It is
easy to check that this group action leaves the set of electron-hole
coherent states invariant (up to phase factors). Note that the
conjugation of  $D(\alpha,\phi)\otimes U(\beta,\varphi)$ with the
nonlinear operator $\mathcal{Z}$ can no longer be written in terms of
two factors acting independently in the two factor spaces. This is
indeed necessary to produce the entanglement of electron-hole coherent
states that we have discussed in Section
\ref{sec:entanglement}.\\
Finally note that group elements that are represented by a simple phase factor in the linear unitary
representation may be less trivial in the nonlinear representation (and vice versa).
For instance $D(0,\phi)\otimes U(0,0)$ acting directly
on any state is just scalar multiplication with $e^{i\phi}$ while
\begin{equation}
  \mathcal{Z}\left[D(0,\phi)\otimes U(0,0)\right] \mathcal{Z}
  =D(0,0) \otimes U(0,\phi)
\end{equation}
happens to be unitary and the angle $\phi$ has swapped from the
$H_3(\mathbb{R})$ to the $SU(2)$ part. Similarly one has
\begin{equation}
  \mathcal{Z}\left[D(0,0)\otimes U(0,\phi)\right] \mathcal{Z}
  =D(0,\phi) \otimes U(0,0)\ .
\end{equation}

\section{Some applications of electron-hole coherent states}
\label{eh-cs-app}

Let us now discuss some applications of electron-hole coherent
states. We will focus on quasi-particle excitations
in a superconductor which can be described in the Hilbert space
$\mathcal{H}_\infty \otimes \mathcal{H}_2$ where the first factor
carries the observables that correspond to space
(here taken to be just the line), and the second factor is spanned by
electron-like excitations represented by the basis state $|+\rangle$ and
hole-like excitations represented by $|-\rangle$.

\subsection{The Bogoliubov-de Gennes equation and
  Heisenberg Dynamics of a Coherent State}
\label{sec:Bdg}

Quasiparticle excitations in
a one-dimensional superconductor are described by the
Bogoliubov-de Gennes equation \cite{deGennes66, tinkham04} which can be written formally as a
Schr\"odinger equation
\begin{equation}
  i \frac{d}{dt} \left| \Psi(t)\right\rangle
  = H \left| \Psi(t)\right\rangle
\end{equation}
with the Hamilton operator
\begin{equation}
  H= \left(\frac{1}{2} P^2 - \mu \right) \sigma_3
  + \Delta(Q) \sigma_1 \ .
\end{equation}
Here, $\mu>0$ is the Fermi energy and $\Delta(q)$ is the real-valued pair
potential (replacing the operator $Q$ by a real eigenvalue $q$) that couples
electron- and hole-like excitations. We have here assumed the simplest setting,
generalizations such as adding a magnetic field or taking a complex-valued pair
potential are straightforward to implement but will not be discussed here.
Note that we assume $\hbar=1$.\\
Let us discuss the dynamics of the Bogoliubov-de Gennes equation
in the Heisenberg picture for either an electron-hole or a product
coherent state.
The Heisenberg equations of motion for the fundamental operators are
easily found to be
\begin{subequations}
  \begin{align}
    \frac{d}{dt} Q_H=&
    P_H\ \sigma_{H,3} \equiv V_H
    \label{dQdt}\\
    \frac{d}{dt} P_H=&- \Delta'(Q_H) \sigma_{H,1}\\
    \frac{d}{dt} V_H= &
    \left(P_H \Delta(Q_H)+ \Delta(Q_H) P_H \right)\sigma_{H,2}
    \\
    \frac{d}{dt} \sigma_1=& -
    2 \left(\frac{1}{2} P_H^2 - \mu \right)\sigma_{H,2} \\
    \frac{d}{dt} \sigma_2=&
    2 \left(\frac{1}{2} P_H^2 - \mu \right)\sigma_{H,1}
    -2 \Delta(Q_H) \sigma_{H,3}
    \\
    \frac{d}{dt} \sigma_3=&
    2 \Delta(Q_H) \sigma_{H,2}
  \end{align}
\end{subequations}
where we have added the index $H$ to refer to the time dependent
operators in the Heisenberg pictures.
We have included here the operator $V_H$ -- equation \eqref{dQdt}
shows the origin of the label 'velocity operator' that we have used before.\\
At time $t=0$ we start the dynamics with the standard representations
of the corresponding operators (i.e $Q_H(0)\equiv Q$ and so on for all
other operators). Now let the system be either in the electron-hole
coherent state $|\alpha \Join \beta\rangle$ or the product coherent
state $|\alpha \otimes \beta\rangle$ with
$\alpha=\frac{q+iv}{\sqrt{2}}$. Both states are localized in
configuration space at $t=0$ near $q$ and we will focus on how the
localization properties change for short time. We will assume here that
any variation of $\Delta(q)$ can be neglected over the extent of the
coherent state and its vicinity and replace it by a constant $\Delta_0$
and neglect all terms $\Delta'(Q)$. Using a short-hand $\langle Q
\rangle_{\Join}\equiv \langle \alpha \Join \beta |Q| \alpha \Join \beta
\rangle$ for electron-hole and $\langle Q\rangle_\otimes \equiv \langle
\alpha \otimes \beta | Q| \alpha \otimes\beta\rangle$ for product
coherent states we then have the following expectation values at $t=0$,
\begin{subequations}
  \begin{align}
    \left.\langle Q \rangle_{\Join}\right|_{t=0}=&
    q, &
    \left.\langle Q \rangle_{\otimes}\right|_{t=0}=&
    q,\\
    \left.\langle P \rangle_{\Join}\right|_{t=0}=&
    v, \frac{1-|\beta|^2}{1+|\beta|^2} &
    \left.\langle P \rangle_{\otimes}\right|_{t=0}=&
    v,\\
    \left.\langle V \rangle_{\Join}\right|_{t=0}=&
    v, &
    \left.\langle V \rangle_{\otimes}\right|_{t=0}=&
    v \frac{1-|\beta|^2}{1+|\beta|^2} .
  \end{align}
\end{subequations}
Their corresponding changes at $t=0$ are
\begin{subequations}
  \begin{align}
    \left.\frac{d}{dt} \langle Q \rangle_{\Join}\right|_{t=0}=&
    v=\left.\langle V \rangle_{\Join}\right|_{t=0}  &
    \left.\frac{d}{dt} \langle Q \rangle_{\otimes}\right|_{t=0}=&
    v \frac{1-|\beta|^2}{1+|\beta|^2}=
    \left.\langle V \rangle_{\otimes}\right|_{t=0}=
    \left.\langle P \rangle_{\otimes}
      \langle \sigma_3 \rangle_{\otimes}
    \right|_{t=0}
    \\
    \left.\frac{d}{dt} \langle P \rangle_{\Join}\right|_{t=0}=&
    0  &
    \left.\frac{d}{dt} \langle P \rangle_{\otimes}\right|_{t=0}=&
    0 \\
    \left.\frac{d}{dt} \langle V \rangle_{\Join}\right|_{t=0}=&
    0  &
    \left.\frac{d}{dt} \langle V \rangle_{\otimes}\right|_{t=0}=&
    2i \Delta_0 v \frac{\beta-\beta*}{1+|\beta|^2}=
    2 \Delta_0
    \left.
      \langle P \rangle_{\otimes}
      \langle \sigma_{2} \rangle_{\otimes}
    \right|_{t=0}\ .
  \end{align}
\end{subequations}
For an electron-hole coherent state this implies that the expectation values
for short time follow a free trajectory with velocity $v$ in the phase space
spanned by position and velocity -- irrespective of the parameter $\beta$ that
describes the quasi-spin. For product coherent states the expectation value of
the position also follows the free motion but the
corresponding velocity can be anything between $-v$ and $v$ depending on
quasi-spin configuration. The structure of electron-hole coherent states thus
seems more adapted to applications in the Bogoliubov-de Gennes equation. This
statement is supported more profoundly by looking at
the variance in position after a short time $\delta t$. For an electron-hole
coherent state this variance is
\begin{subequations}
  \begin{align}
  \mathrm{Var}_{\Join} \left[Q_H(0)\right]=&
  \frac{1}{2} &
  \mathrm{Var}_\otimes \left[Q_H(0)\right]=&
  \frac{1}{2}\\
  \mathrm{Var}_{\Join} \left[ Q_H(\delta t)=Q+\delta t V \right]=&
  \frac{1}{2} \left(1 + \delta t^2 \right)&
  \mathrm{Var}_\otimes \left[ Q_H(\delta t) \right]=&
  \frac{1}{2}\left(1 + \left(1+v^2\frac{8|\beta|^2}{(1+|\beta|^2)^2}\right)
    \delta t^2
  \right)\ .
  \end{align}
\end{subequations}
Thus product coherent states generally disperse much quicker than
electron-hole coherent states. The latter follow the short-time
dispersion for a free scalar wave packet. The origin of the quick
dispersion is obvious, the electron-like component of the product
coherent state moves with velocity $v$ while the hole-like component
moves with opposite velocity $-v$ and such that the total quasi-spinor
wave packet breaks up. This does not happen for the electron-hole
coherent state, where both components have the same velocity $v$.\\
The time scale connected to quasi-spin rotations is $\sim 1/\Delta_0$.
One may expect that this is the time scale where the electron-hole
coherent state may change its character and, for instance, ceases to be
well localised in the position-velocity phase space. A more detailed
analysis of the wave packet dynamics on intermediate and longer time
scales in the Schr\"odinger picture reveals an interesting interplay
between space and spinor evolutions for electron-hole coherent states
where the velocity $v$ is close to the Fermi velocity $v_F=\sqrt{2
\mu}$ (or momentum close to the Fermi momentum $p_F=\sqrt{2\mu}$ which
is equivalent to the Fermi velocity in value as we have used units such
that the mass is unity) \cite{GnLL}.

\subsection{Electron-hole coherent state representation}
\label{sec:eh-cs-representation}

The resolution of unity \eqref{resolution_unity_eh} implies that we
can use electron-hole coherent states to represent states in $\mathcal{H}_\otimes$
as functions of two complex parameters $\alpha$ and $\beta$
\begin{equation}
  |\Psi\rangle \mapsto \Psi(\alpha,\alpha^*,\beta,\beta^*)=
  \langle \alpha \Join \beta |\Psi\rangle\ .
\end{equation}
The Hilbert space spanned by such functions $\Psi(\alpha,\alpha^*,\beta,\beta^*)$
under the above map consists of functions of the form
\begin{equation}
  \Psi(\alpha,\alpha^*,\beta,\beta^*)=
  \frac{e^{-|\alpha|^2/2}}{1+|\beta|^2}\left( u_{|\Psi\rangle}(\alpha^*) + \beta v_{|\Psi\rangle}(\alpha) \right)
\end{equation}
where $u_{|\Psi\rangle}(\alpha^*)$ is analytic in $\alpha^*$,
$v_{|\Psi\rangle}(\alpha)$ analytic in $\alpha$ and both integrals
$\int_{\mathbb{C}} d^2\alpha |u_{|\Psi\rangle}(\alpha^*)|^2
e^{-|\alpha|^2}$, and $\int_{\mathbb{C}} d^2\alpha
|v_{|\Psi\rangle}(\alpha)|^2 e^{-|\alpha|^2}$ exist. The scalar product
in this representation is
\begin{equation}
 \langle \Psi_1|\Psi_2\rangle= \frac{2}{\pi^2}
 \int d^2\alpha d^\beta
 \frac{1}{(1+|\beta|^2)^2}
 \Psi_1(\alpha,\alpha^*,\beta,\beta^*)^*
 \Psi_2(\alpha,\alpha^*,\beta,\beta^*).
\end{equation}
It is more convenient to use the functions
\begin{equation}
  f_{|\Psi\rangle}(\alpha,\alpha^*,\beta)\equiv
  u_{|\Psi\rangle}(\alpha^*) + \beta v_{|\Psi\rangle}(\alpha)
\end{equation}
with a rescaled scalar product
\begin{equation}
  \langle \Psi_1|\Psi_2\rangle= \frac{2}{\pi^2}
  \int d^2\alpha d^2\beta\
  \frac{e^{-|\alpha|^2}}{(1+|\beta|^2)^3}
  f_{\Psi_1}^*
  f_{\Psi_2}
  = \frac{1}{\pi} \int d^2\alpha\ \left(
    u_{|\Psi_1\rangle}(\alpha^*)^*  u_{|\Psi_2\rangle}(\alpha^*)
    +
    v_{|\Psi_1\rangle}(\alpha)^*  v_{|\Psi_2\rangle}(\alpha)
  \right)\ .
\end{equation}
Using
\begin{subequations}
  \begin{align}
    a  \left|\alpha\Join \beta)\right)=&
    \left(
      \alpha \left(1-\beta^* \frac{\partial}{\partial \beta^*}\right) +
      \alpha^* \beta^*\frac{\partial}{\partial \beta^*}
    \right) \left|\alpha\Join \beta\right) \\
    a^\dagger\left|\alpha\Join \beta\right)=&
    \left(
      \frac{\partial}{\partial \alpha} +
      \frac{\partial}{\partial \alpha^*}
    \right) \left|\alpha \Join \beta\right)
    \\
    \sigma_1 \left|\alpha \Join \beta\right)=&
    \left(\beta^*\left(1-\beta^*\frac{\partial}{\partial \beta^*}\right)+
      \frac{\partial}{\partial \beta^*}\right)
    \left|\alpha^*\Join \beta\right)=
    \left(\beta^*\left(1-\beta \frac{\partial}{\partial \beta}\right)+
      \frac{\partial}{\partial \beta}\right)
    \left|\alpha^*\Join \beta^*\right)\\
    \sigma_2 \left|\alpha\Join \beta\right)=&
    \left(-i \beta^*\left(1-\beta^*\frac{\partial}{\partial \beta^*}\right)+
      i \frac{\partial}{\partial \beta^*}\right)
    \left|\alpha^* \Join \beta\right)=
    \left(-i \beta^*\left(1-\beta \frac{\partial}{\partial \beta}\right)+
      i \frac{\partial}{\partial \beta}\right)
    \left|\alpha^*\Join \beta^*\right)
    \\
    \sigma_3 \left|\alpha \Join \beta\right)=&
    \left(
      (1-\beta^*\frac{\partial}{\partial \beta^*})
      -
      \beta^*\frac{\partial}{\partial \beta^*}
    \right)
    \left|
      \alpha\Join \beta
    \right)
  \end{align}
  \label{eq:operator_action}
\end{subequations}
one may then rewrite the Bogoliubov-de Gennes equation as
a nonlocal partial differential equation
\begin{equation}
  \begin{split}
    i \frac{\partial }{\partial t}
    f_{|\Psi\rangle}(\alpha,\alpha^*,\beta)=&
    \frac{1}{4}\left[
      \left(1-\beta\frac{\partial}{\partial
          \beta}\right)\left(1-{\alpha^*}^2+2\alpha^*\frac{\partial}{\partial
          \alpha^*}-\frac{\partial^2}{\partial {\alpha^*}^2}\right) \right]f_{|\Psi\rangle}(\alpha,\alpha^*,\beta)\\
    &\quad-\left[\beta\frac{\partial}{\partial
        \beta}\left(1-{\alpha}^2+2\alpha\frac{\partial}{\partial
          \alpha}-\frac{\partial^2}{\partial {\alpha}^2}\right)
    \right]f_{|\Psi\rangle}(\alpha,\alpha^*,\beta)\\
    &\quad-\mu\left[1-2\beta\frac{\partial}{\partial \beta}\right]f_{|\Psi\rangle}(\alpha,\alpha^*,\beta)\\
    &\quad+ \Delta
    \left[\beta\left(1-\beta \frac{\partial}{\partial \beta}\right) +\frac{\partial}{\partial \beta}\right]
    f_{|\Psi\rangle}(\alpha^*,\alpha,\beta)
  \end{split}
  \label{BdG_csrep}
\end{equation}
for a scalar (complex-valued) function
on phase space (with constant pair potential $\Delta(Q)=\Delta_0$).
The nonlocal character of \eqref{BdG_csrep} can be seen from the
different order of arguments in the last term (where $\alpha$ and
$\alpha^*$ have been interchanged).
\\
This representation in terms of functions
$f_{|\Psi\rangle}$ has similarities to the well-known Bargmann
representations -- with the drawback that $f_{|\Psi\rangle}$
is not an analytic function of $\alpha$.

\subsection{The generalized Husimi function and its limitations}
\label{sec:phasespace-rep}

\begin{figure}[ht]
  \includegraphics[width=0.45\textwidth]{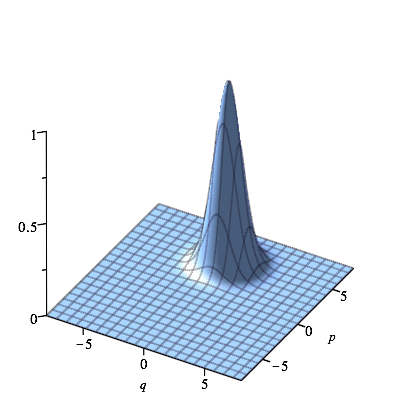}\hfill
  \includegraphics[width=0.45\textwidth]{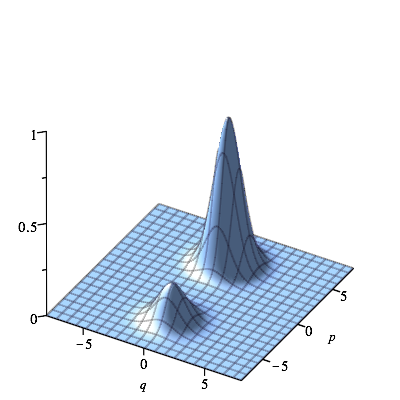}\\
  \includegraphics[width=0.45\textwidth]{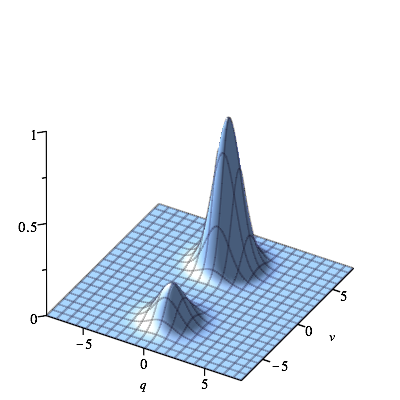}\hfill
  \includegraphics[width=0.45\textwidth]{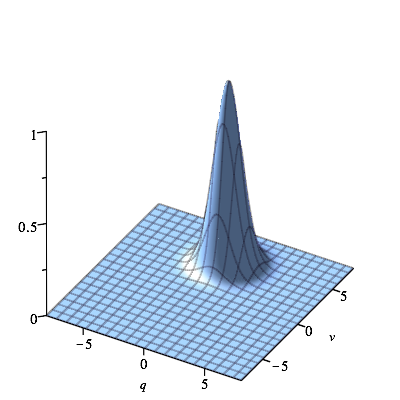}
  \caption{Reduced phase space functions
    $h_{\otimes,\rho}^{(\mathrm{red})}(\alpha,\alpha^*)$ (top figures
    with $\alpha=(q+ip)/\sqrt{2}$)
    and $h_{\Join,\rho}^{(\mathrm{red})}(\alpha,\alpha^*)$ (bottom
    figures with $\alpha=(q+iv)/\sqrt{2}$) for product coherent state $\rho=|\alpha_0 \otimes
    \beta_0\rangle \langle \alpha_0 \otimes \beta_0|$ (left figures)
    and an electron-hole coherent state $\rho=|\alpha_0 \Join
    \beta_0\rangle \langle \alpha_0 \Join \beta_0|$
    (with $\alpha_0=4 i /\sqrt{2}$ and $\beta_0=1/2$). The top left
    (top right)
    figure is identical to the bottom right (bottom left) figure apart
    from the axis description which changes from momentum $p$ to
    velocity $v$.
  }
  \label{fig1}
\end{figure}

\begin{figure}[h]
  \includegraphics[width=0.33\textwidth]{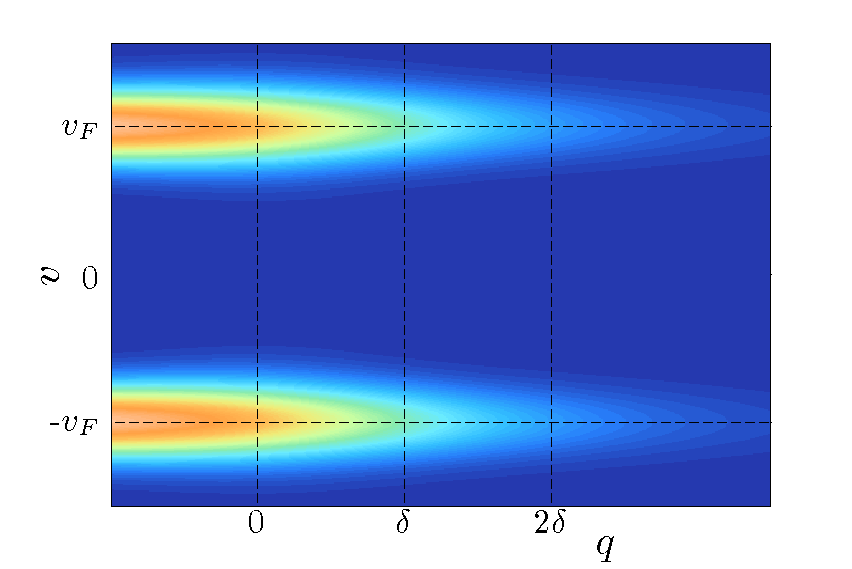}
  \includegraphics[width=0.33\textwidth]{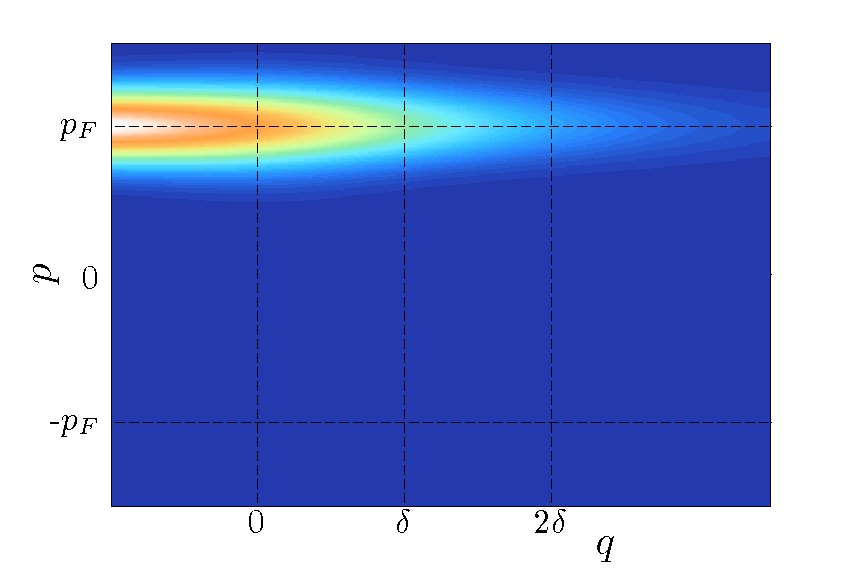}
  \hfill
  \includegraphics[width=0.29\textwidth]{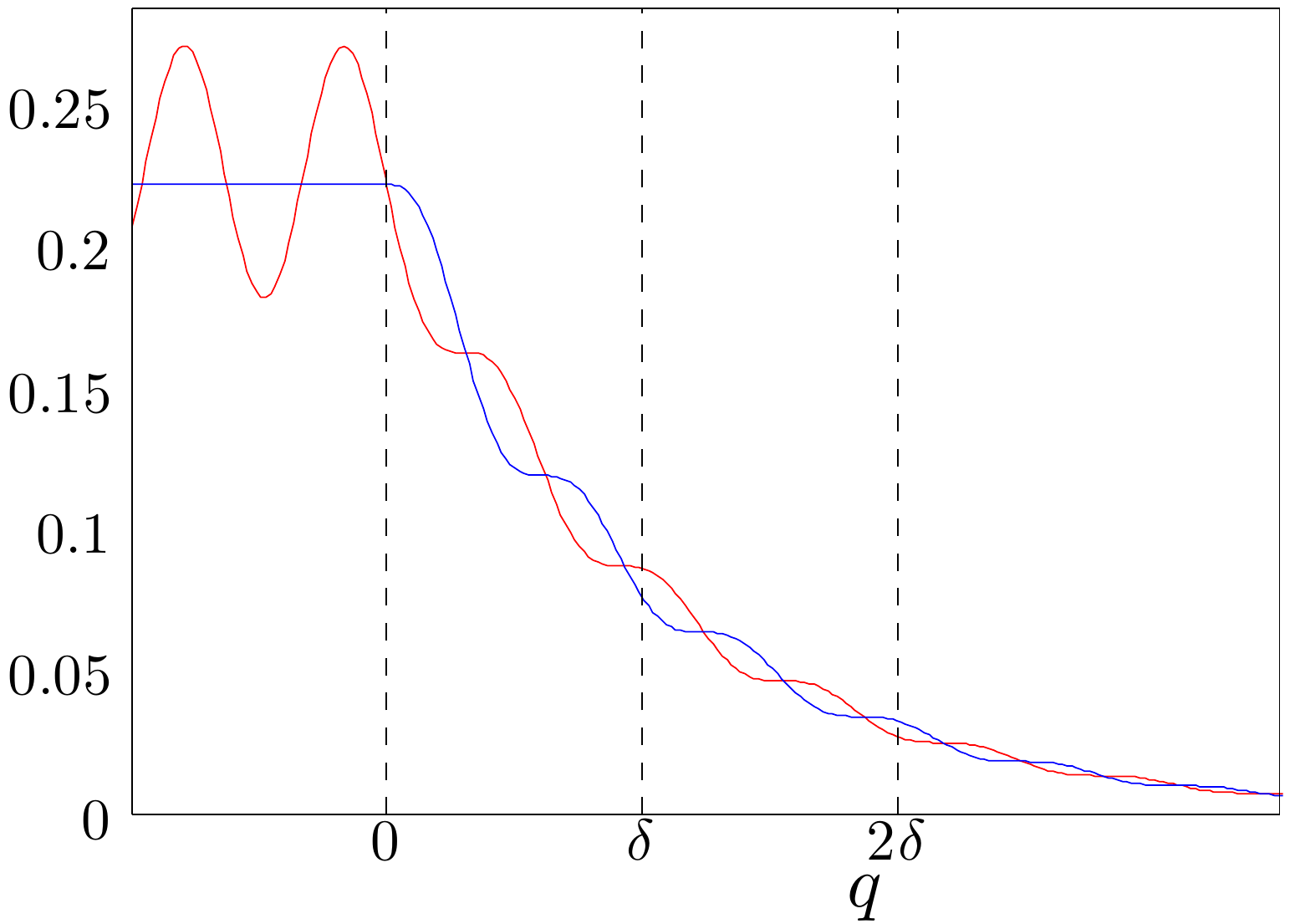}\\
  \includegraphics[width=0.33\textwidth]{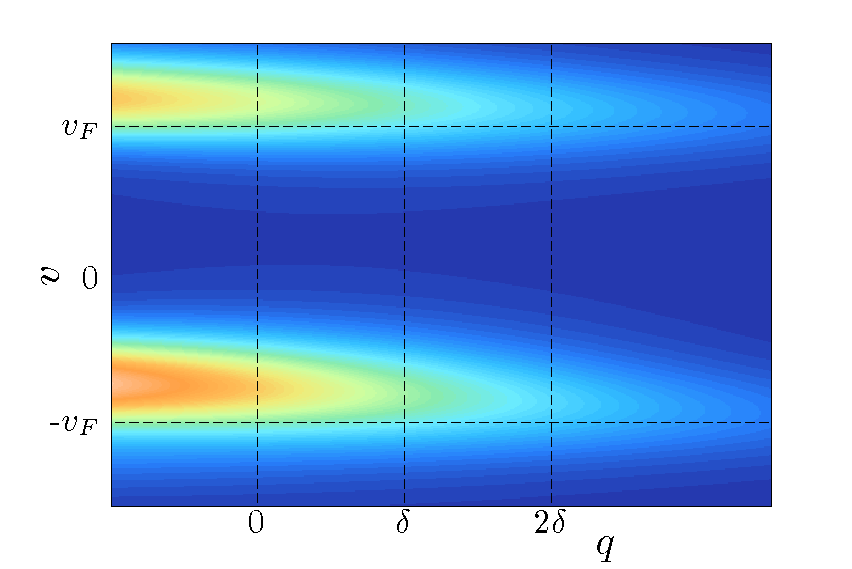}
  \includegraphics[width=0.33\textwidth]{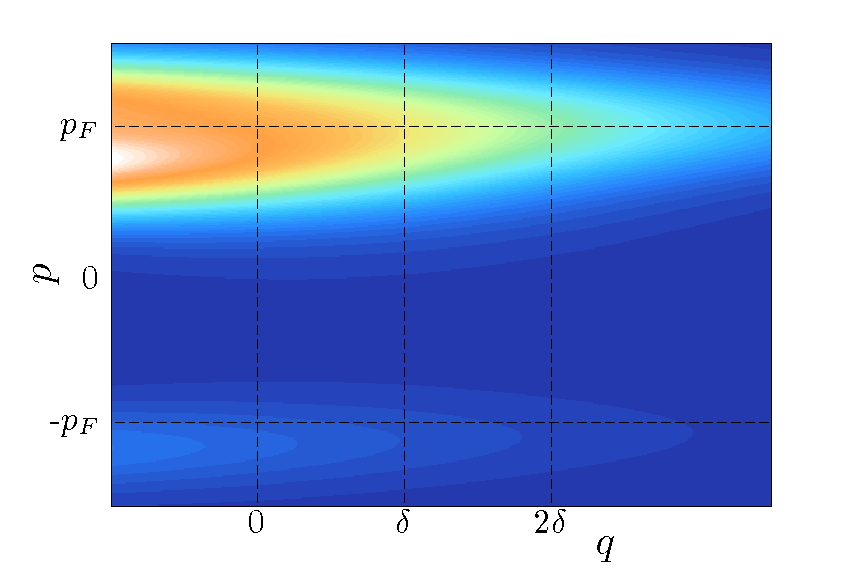}
  \hfill
  \includegraphics[width=0.29\textwidth]{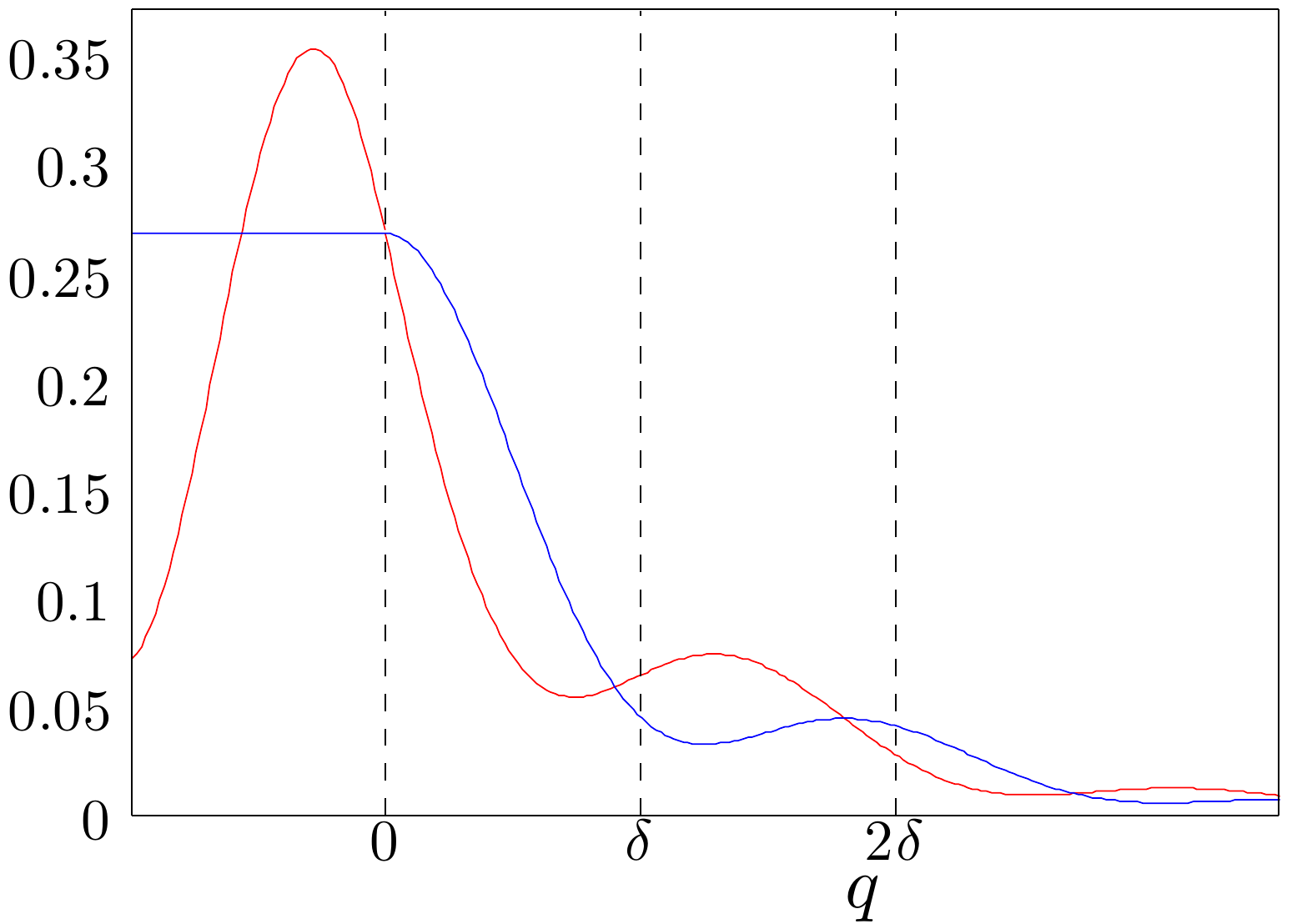}\\
  \includegraphics[width=0.33\textwidth]{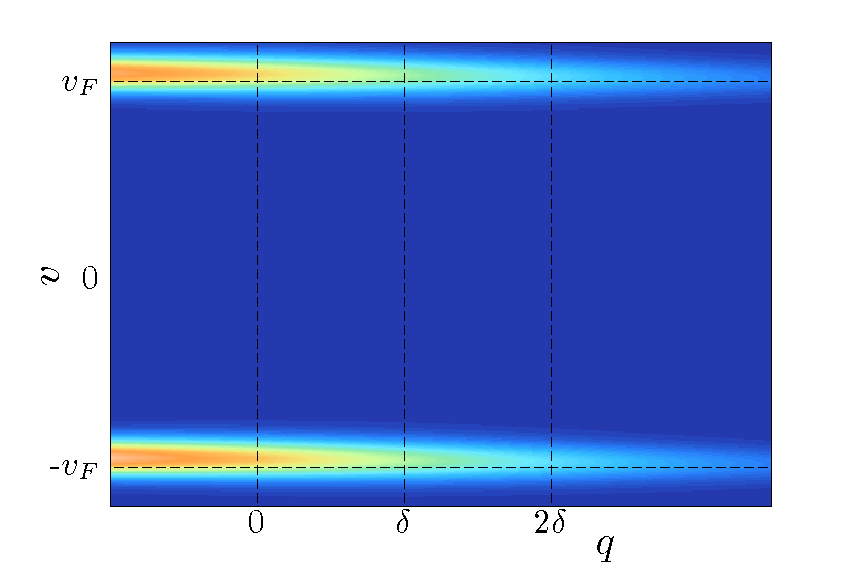}
  \includegraphics[width=0.33\textwidth]{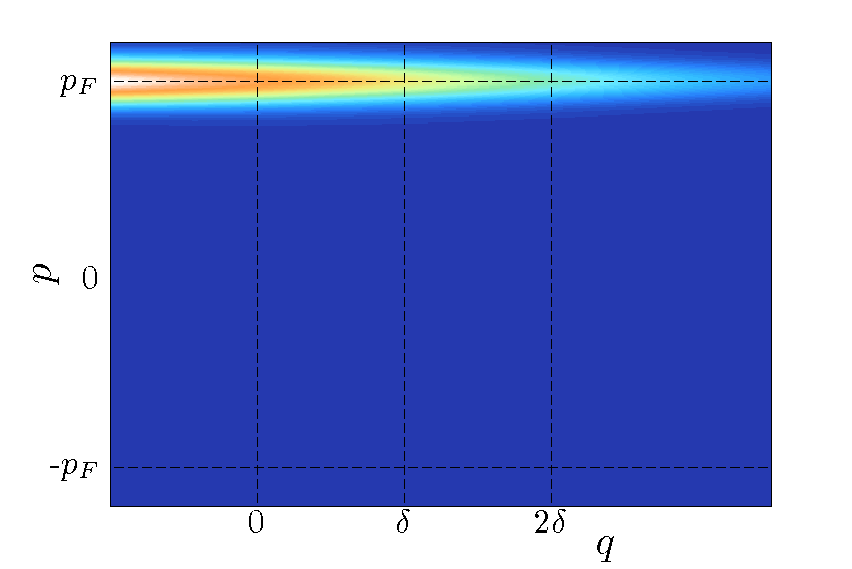}
  \hfill
  \includegraphics[width=0.29\textwidth]{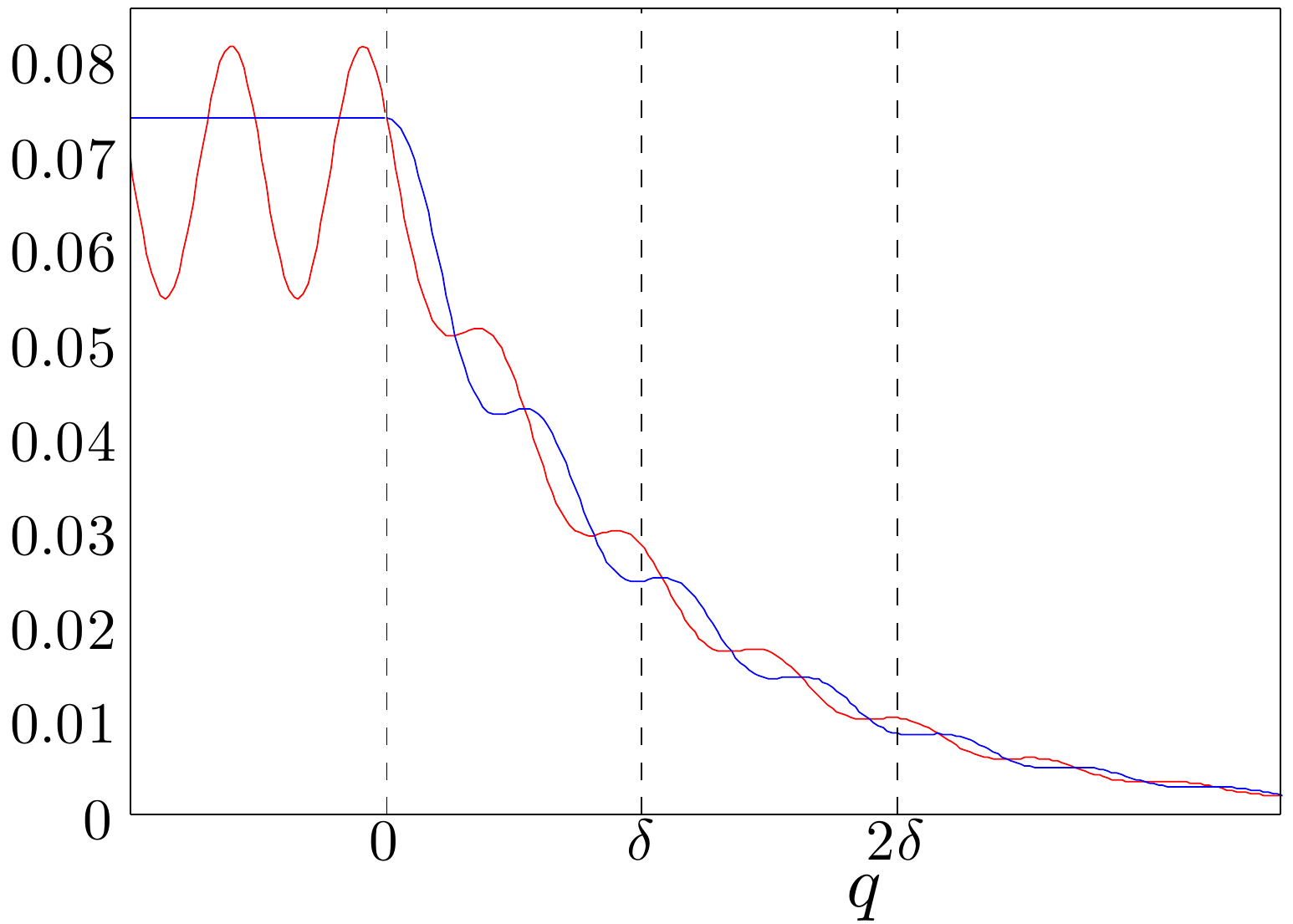}
  \caption{Andreev reflection from a pair potential
    $\Delta(q)= \Delta_0 \theta(q)$.  Stationary scattering states at energy $E$
    for an incoming electron wave are shown in
    terms of the reduced position-velocity phase space function
    $h_{\Join,\rho}^{(\mathrm{red})}(\alpha,\alpha^*)$ (left column),
     the reduced position-momentum phase space function
    $h_{\otimes,\rho}^{(\mathrm{red})}(\alpha,\alpha^*)$ (middle column),
    and their spinor wave functions (right column, red: absolute
    value squared of electron component, blue: absolute value squared
    of hole component). The scale $\delta$ is the penetration depth defined in \eqref{penetrationdepth}. \\
    Upper row: $E=0$, $\Delta_0=2$, $\mu=10$;
    middle row: $E=5$, $\Delta_0=8$, $\mu=10$;
    middle row: $E=10$, $\Delta_0=20$, $\mu=100$.
  }
  \label{fig2}
\end{figure}

\begin{figure}[h]
  \includegraphics[width=0.33\textwidth]{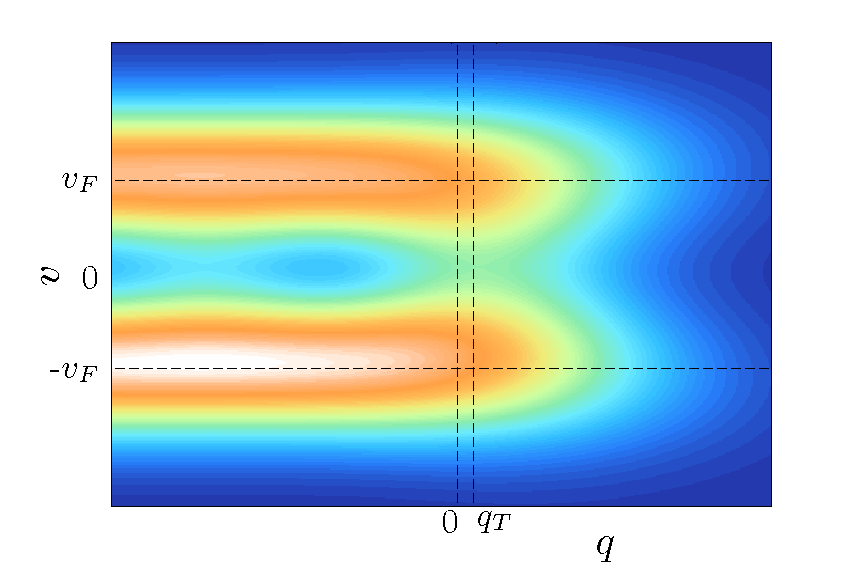}
  \includegraphics[width=0.33\textwidth]{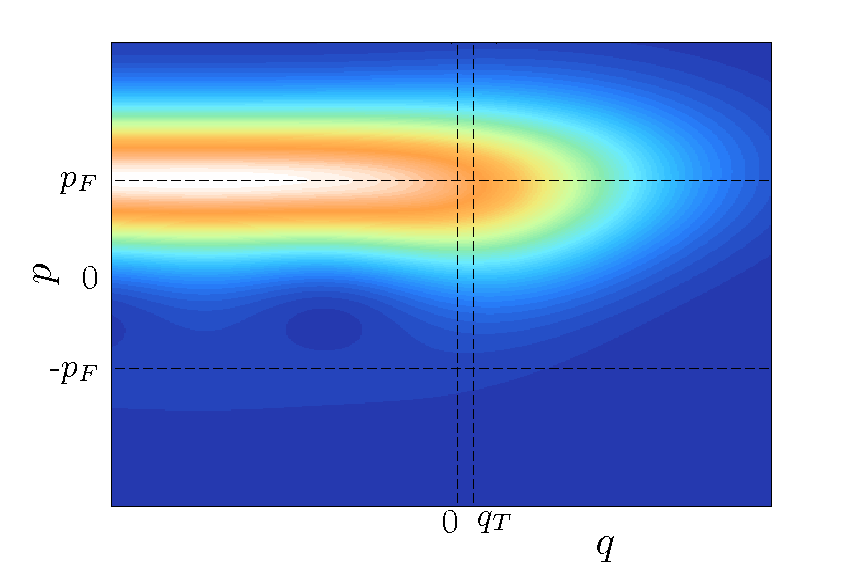}
  \hfill
  \includegraphics[width=0.29\textwidth]{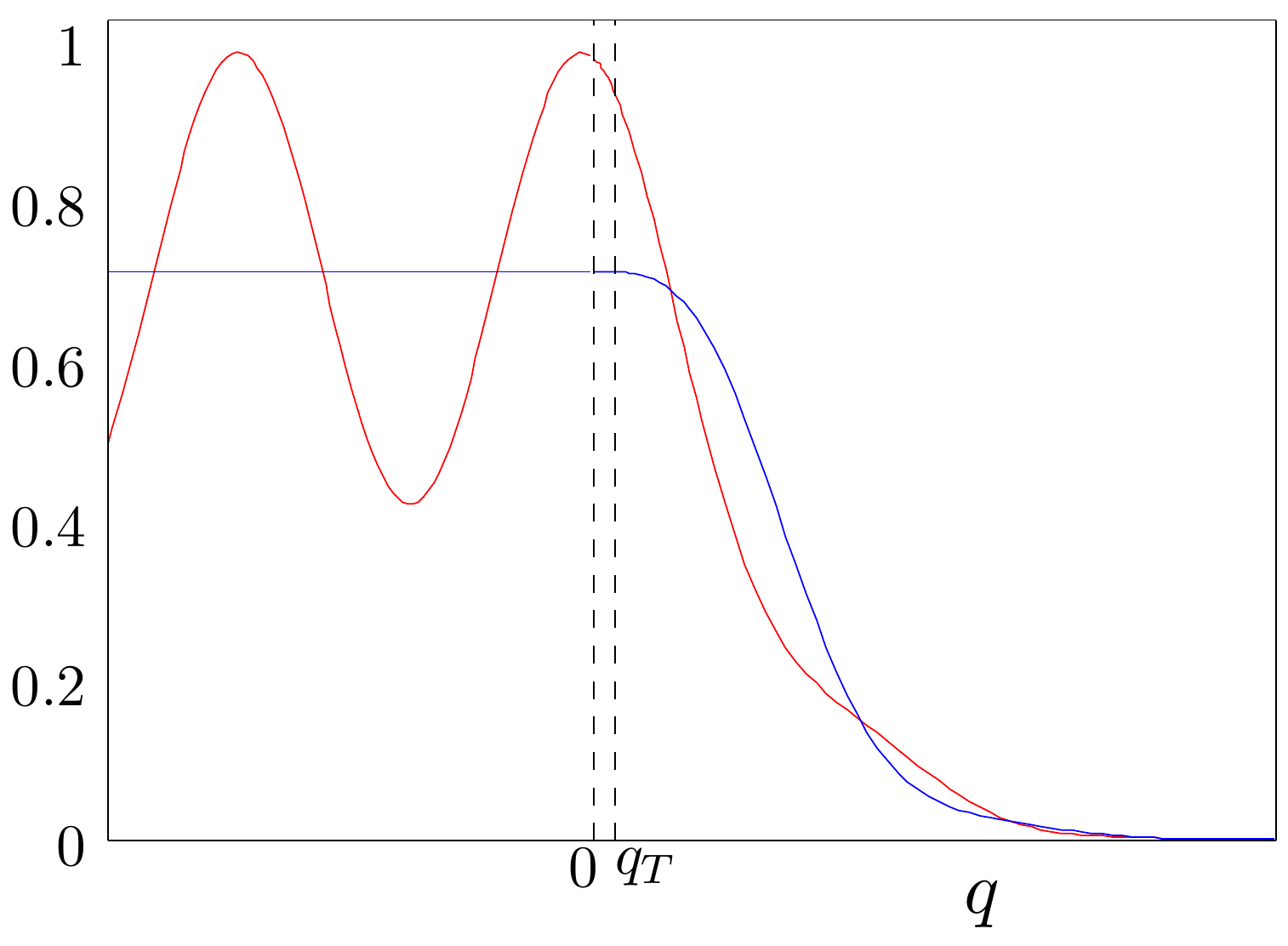}\\
  \includegraphics[width=0.33\textwidth]{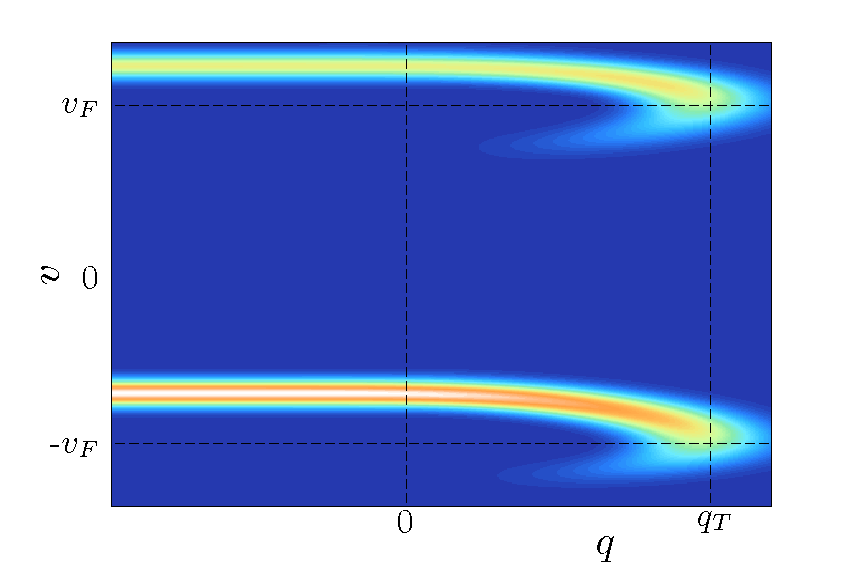}
  \includegraphics[width=0.33\textwidth]{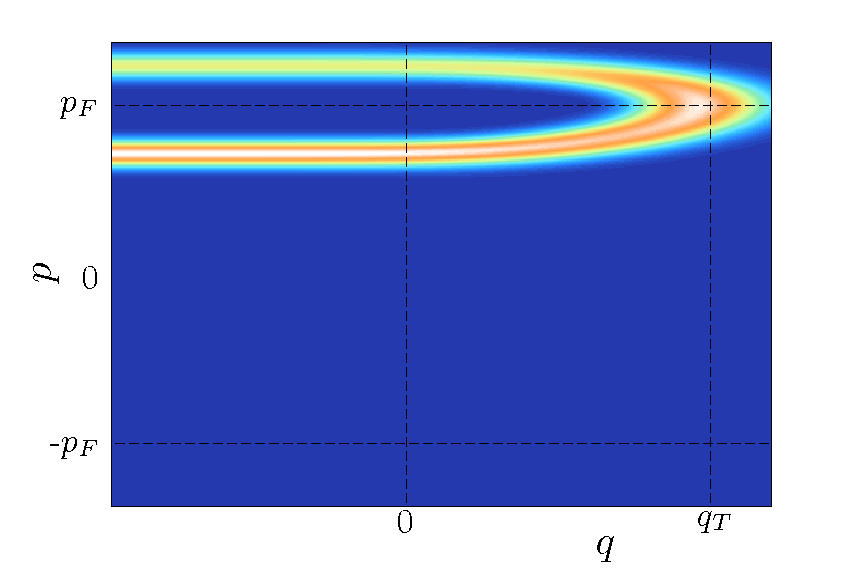}
  \hfill
  \includegraphics[width=0.29\textwidth]{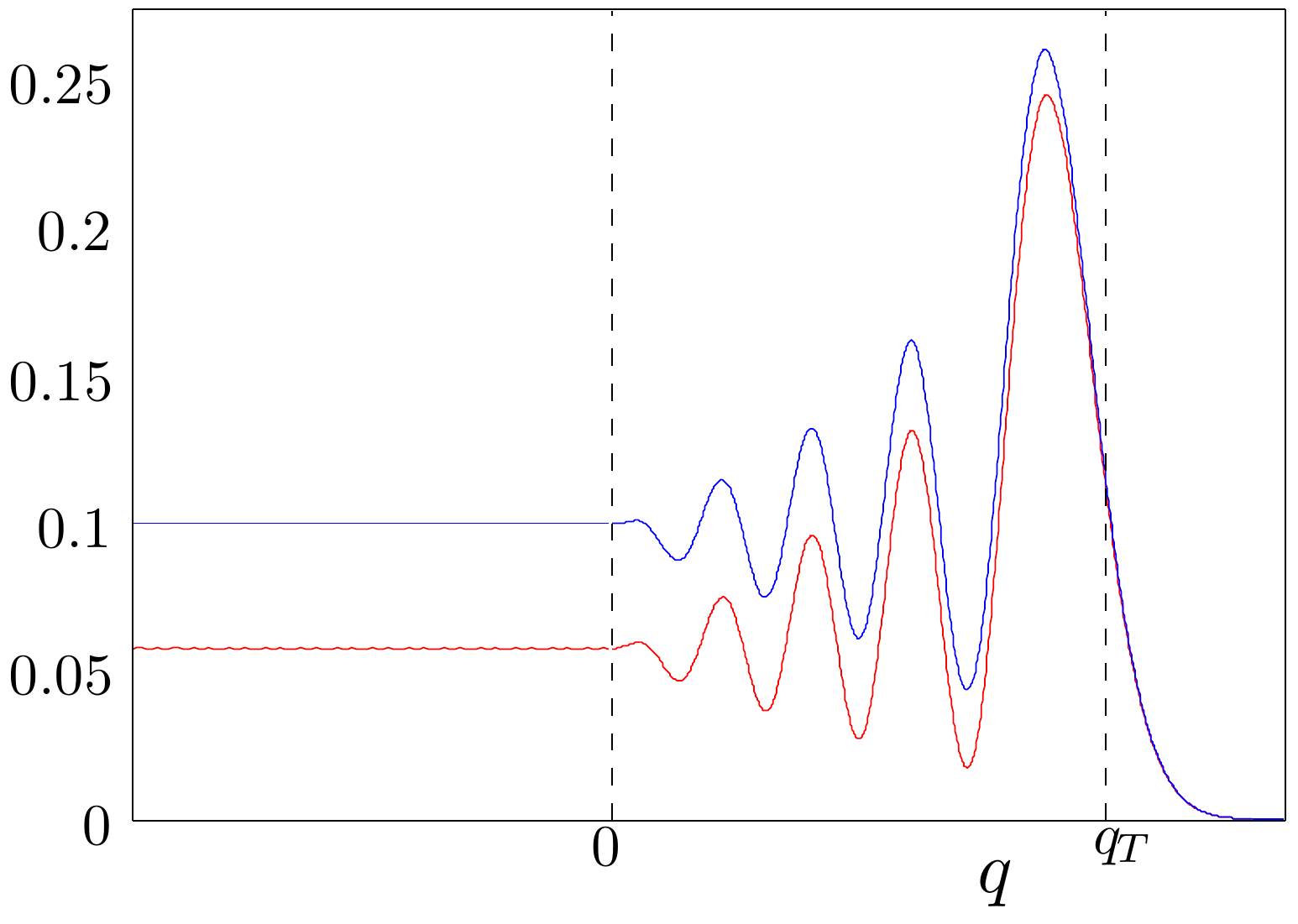}\\
  \includegraphics[width=0.33\textwidth]{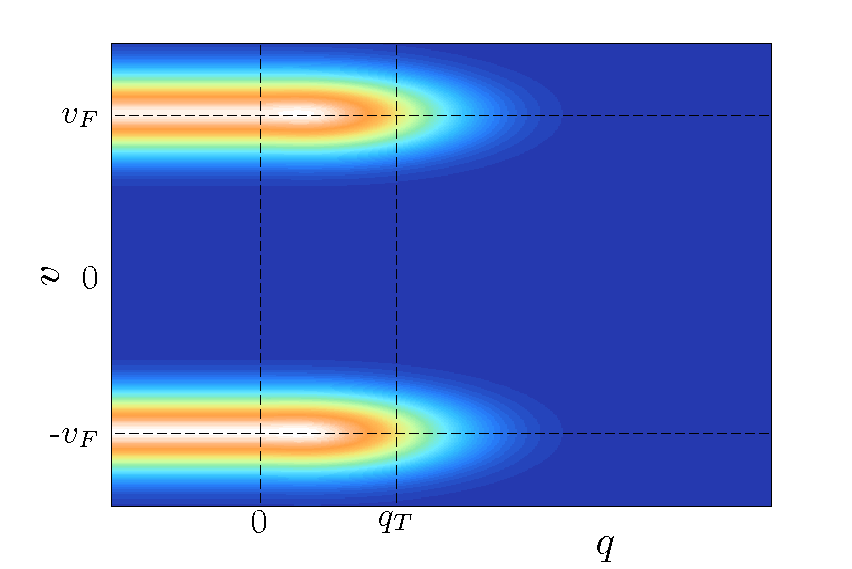}
  \includegraphics[width=0.33\textwidth]{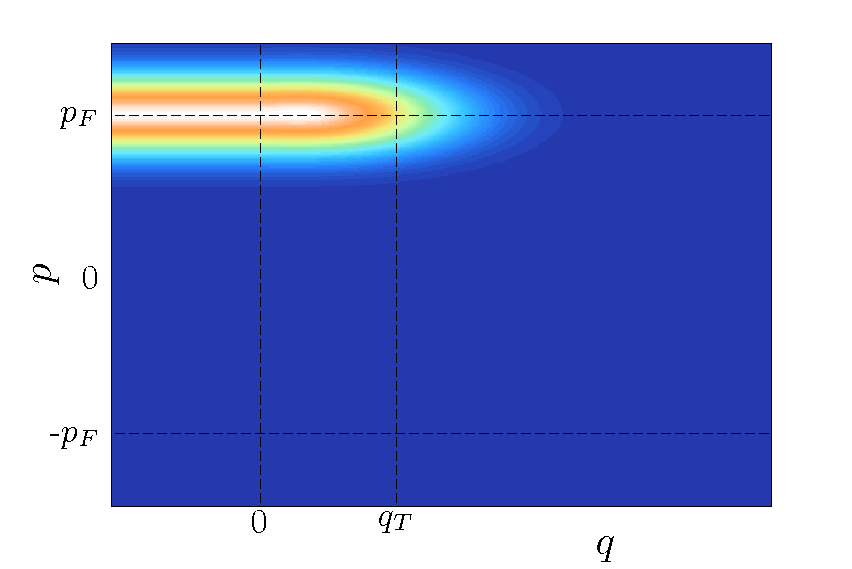}
  \hfill
  \includegraphics[width=0.29\textwidth]{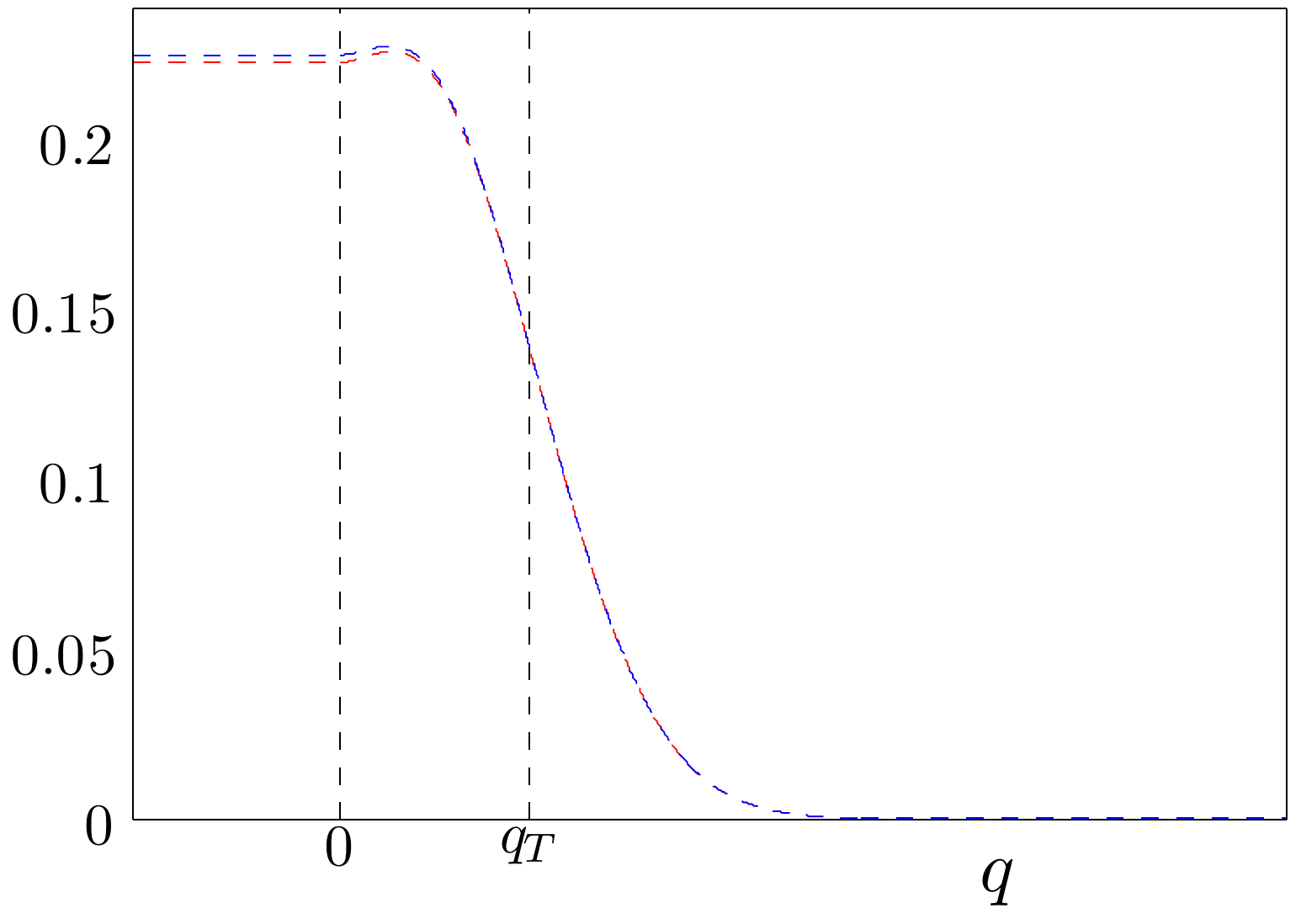}\\
  \includegraphics[width=0.33\textwidth]{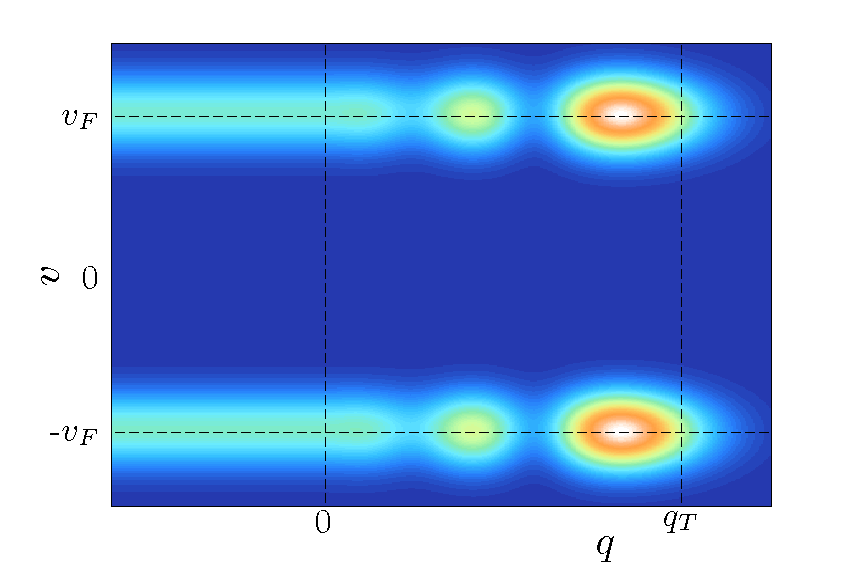}
  \includegraphics[width=0.33\textwidth]{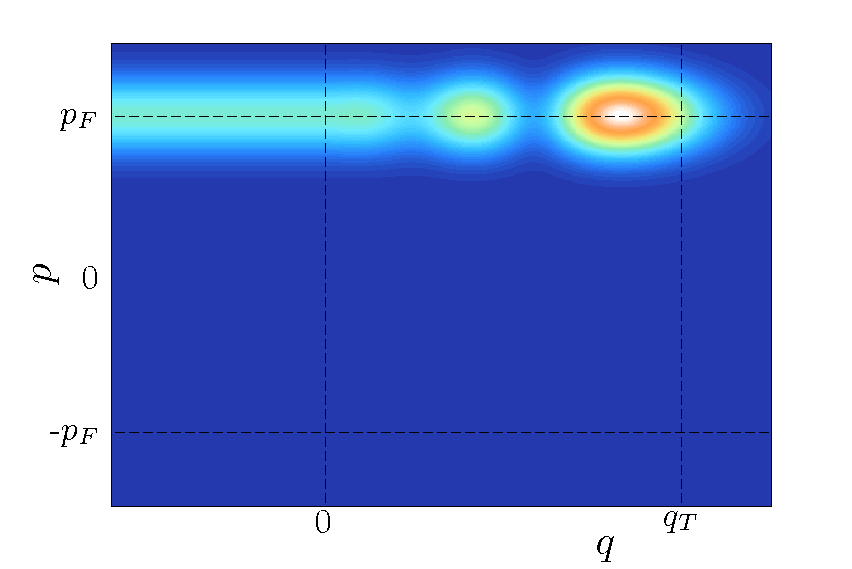}
  \hfill
  \includegraphics[width=0.29\textwidth]{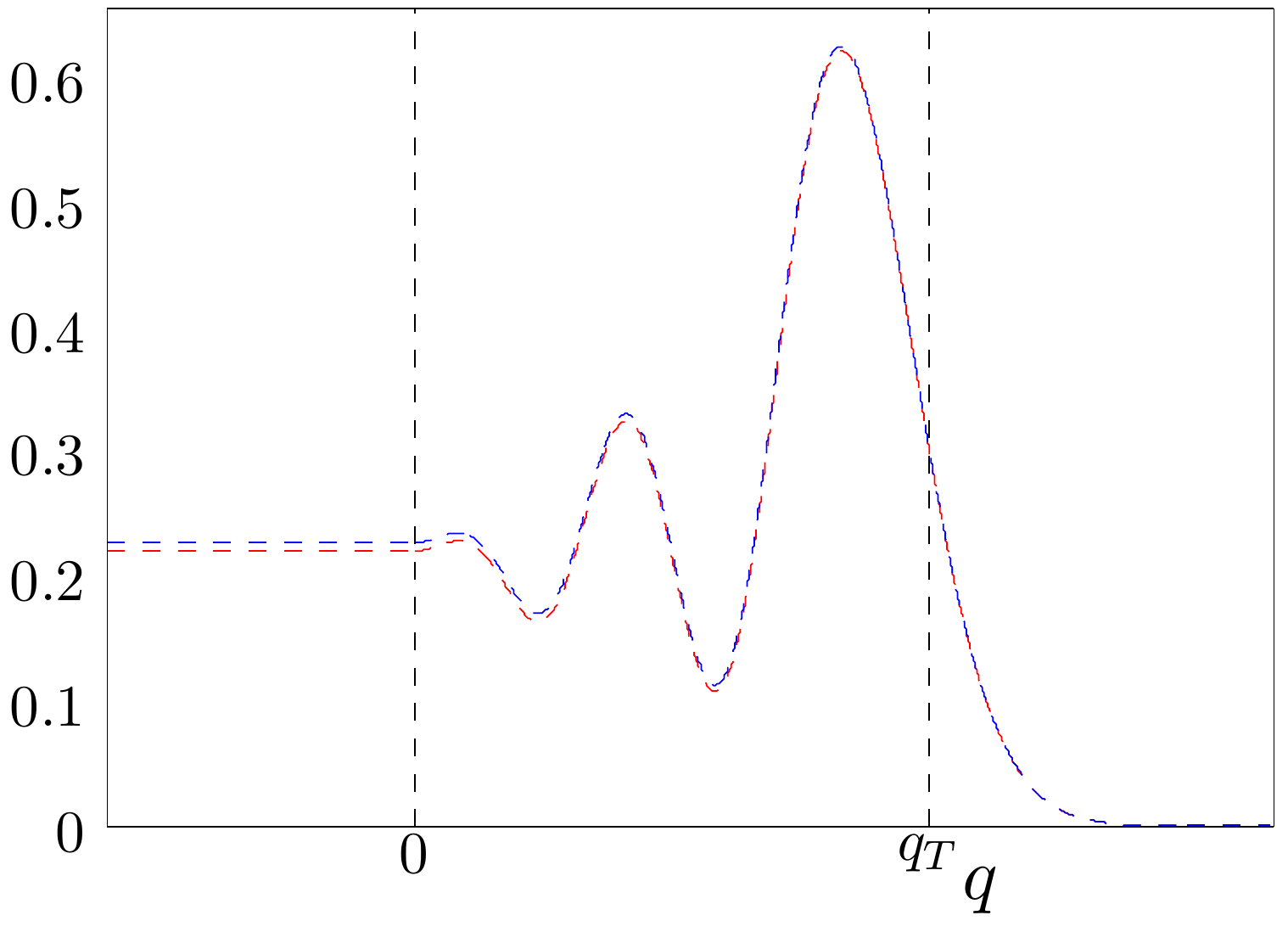}
  \caption{Andreev reflection from a pair potential
    $\Delta(q)=\nu x \theta(x)$. Stationary scattering states at energy $E$
    for an incoming electron wave are shown in terms of the reduced position-velocity phase space function
    $h_{\Join,\rho}^{(\mathrm{red})}(\alpha,\alpha^*)$ (left column), the reduced position-momentum phase space function
    $h_{\otimes,\rho}^{(\mathrm{red})}(\alpha,\alpha^*)$ (middle column),
    and their spinor wave functions (right column, red: absolute
    value squared of electron component, blue: absolute value squared
    of hole component).
    The scale is given in terms of the turning point $q_T=\nu/E$.\\
    First row: $E=0.1$, $\mu=1$, $\nu=0.75$; second row: $E=50$, $\mu=100$, $\nu=10$;
    third row: $E=0.1$, $\mu=10$, $\nu=0.002$; fourth row: $E=0.3$, $\mu=10$, $\nu=0.002$.
  }
  \label{fig3}
\end{figure}

The standard coherent states $\{|\alpha\rangle\}$ allow constructing a
phase space representations of quantum mechanics, where a state of the
system $\rho$ (mixed or pure) is represented by the Husimi function
\begin{equation}
  h_{\rho} (\alpha,\alpha^*)= \langle\alpha| \rho|\alpha\rangle
\end{equation}
which has all formal properties of a classical probability density function
(non-negative and normalised). Analogously an operator $A$ is represented by
its symbol
\begin{equation}
  \mathcal{A} (\alpha,\alpha^*)= \langle\alpha| A|\alpha\rangle \ .
\end{equation}
There is a one-to-one correspondence between operators and their
symbols, or quantum states and their Husimi functions. The dynamics of
the state and evaluation of any expectation values can be expressed
entirely in terms of symbols and the Husimi function (by adding some
further structure such as a non-commutative product for functions on
the phase space). Many generalizations of coherent states (e.g. the
$SU(2)$ coherent states) allow similar constructions. On the tensor
product space $\mathcal{H}_\infty \otimes \mathcal{H}_2$ the product
coherent states lead to an analogous one-to-one correspondence between
operators on $\mathcal{H}_\infty \otimes \mathcal{H}_2$ and their
functions on the phase space $\mathbb{C}^2 \times S^2$. For an operator
$A$ the corresponding function is again known as the symbol of $A$ and
is defined as
\begin{equation}
  \mathcal{A}_\otimes(\alpha,\alpha^*,\beta,\beta^*)=
  \langle \alpha \otimes \beta| A|\alpha \otimes \beta\rangle.
\end{equation}
If $\rho \in \mathcal{H}_\infty \otimes \mathcal{H}_2$ is a (possibly
mixed) state of the system then the corresponding Husimi function
defined via product coherent states is
\begin{equation}
  h_{\otimes,\rho}(\alpha,\alpha^*,\beta,\beta^*)
  =
  \left\langle \alpha \otimes \beta\right| \rho\left|\alpha \otimes \beta\right\rangle.
\end{equation}

If one tries to repeat these constructions with
electron-hole coherent states one finds that there
is \emph{no one-to-one correspondence} between
operators and the corresponding `symbols' in electron-hole coherent
states.
For instance the two Hermitian
operators $\sigma_1 \frac{{a^\dagger}^2 +a^2}{2}$ and
$\sigma_1a^\dagger a$ would have the same symbol
\begin{equation}
  \left\langle \alpha \Join \beta \right| \sigma_1a^\dagger
  a\left|\alpha \Join \beta\right\rangle
  =\frac{\beta^* {\alpha^*}^2e^{{\alpha^*}^2}+\beta\alpha^2e^{\alpha^2}}{1+|\beta|^2}e^{-|\alpha|^2}
  =
  \left\langle \alpha \Join \beta \right| \sigma_1 \frac{a^2+{a^\dagger}^2}{2}\left|\alpha \Join \beta\right\rangle\
\end{equation}
and this also equals the would-be symbol of the non-Hermitian
operators
$\sigma_1 a^2$ and $\sigma_1 {a^\dagger}^2$.\\
Analogously the `Husimi function'
\begin{equation}
  h_{\Join,\rho}(\alpha,\alpha^*,\beta,\beta^*)
  =
  \left\langle \alpha \Join \beta\right| \rho\left|\alpha \Join \beta\right\rangle
\end{equation}
does not give a complete description of the state. It can still be
useful in order to analyse visually how a given state is distributed in
the position-velocity phase space. One may compare this to the
distribution in the position-momentum phase space \emph{via} the
corresponding Husimi function
$h_{\otimes,\rho}(\alpha,\alpha^*,\beta,\beta^*)$  in product coherent
states (giving a complete description). For this one may use the
reduced phase space functions
\begin{align}
  h_{\otimes,\rho}^{(\mathrm{red})}(\alpha,\alpha^*)=&\frac{2}{\pi}
                                                       \int \frac{d^2\beta}{(1+|\beta|^2)^2}  h_{\otimes,\rho}(\alpha,\alpha^*,\beta,\beta^*),\\
  h_{\Join,\rho}^{(\mathrm{red})}(\alpha,\alpha^*)=&\frac{2}{\pi}
                                                     \int \frac{d^2\beta}{(1+|\beta|^2)^2}\  h_{\Join,\rho}(\alpha,\alpha^*,\beta,\beta^*),
\end{align}
where the quasi-spin variable is integrated out.

Figure \ref{fig1} shows such reduced phase space distributions if the
state is in a particular electron-hole coherent state $\rho_{\otimes}=
| \alpha_0 \otimes \beta_0 \rangle \langle \alpha_0 \otimes \beta_0|$
or a particular product coherent state $\rho_{\Join}= |\alpha_0 \Join
\beta_0 \rangle \langle \alpha_0 \Join \beta_0 |$.  The quasi-spin
position of the product state $\rho_\otimes$ cannot be seen in the
reduced function $h_{\otimes,\rho}^{(\mathrm{red})}(\alpha,\alpha^*)$
in position-momentum phase space. Comparing it to the reduced function
$h_{\Join,\rho}^{(\mathrm{red})}(\alpha,\alpha^*)$ in position-velocity
phase space reveals that the original state has both an electron and a
hole component. For an electron-hole coherent state $\rho_{\Join}$ the
roles of the two reduced phase space functions is interchanged.

An interesting physical phenomenon that may be analysed visually in the
phase space is the Andreev reflection at a boundary between a normal
conducting region (where the pair potential vanishes $\Delta(q)=0$) and
a superconducting region ($\Delta(q)\neq 0$). An incoming electron-like
state with energy $E$  close to the Fermi energy $\mu$  is then
reflected as a hole-like state with (almost) opposite velocity while
the momentum has hardly changed.

Figure \ref{fig2} shows reduced  phase space functions for the
stationary scattering states at energy $E$ from a staircase pair
potential $\Delta(q)= \Delta_0 \theta(q)$ together with the
corresponding spinor wave function. For $q>0$ the (envelope of the)
intensity of the spinor wave function decays exponentially $\propto
e^{-q/\delta}$ where
\begin{equation}
  \delta^{-1}=2 \mathrm{Im} \sqrt{2(\mu+i\sqrt{\Delta_0^2-E^2})}.
  \label{penetrationdepth}
\end{equation}
For these states the reduced representation
$h_{\Join,\rho}^{(\mathrm{red})}(\alpha,\alpha^*)$ in the
position-velocity phase space separates the incoming electron and the
reflected hole amplitudes and gives a more detailed picture of the
dynamics. However this does not necessarily imply that the
electron-hole coherent state representation should always be used on
its own. For instance in the middle row of Figure \ref{fig2} parameters
have been chosen such that the incoming electron has a velocity that is
somewhat above the Fermi velocity and the velocity of the reflected
hole is (in absolute value) somewhat below the Fermi velocity. This is
well resolved in the position-velocity phase space function
$h_{\Join,\rho}^{(\mathrm{red})}(\alpha,\alpha^*)$ built from
electron-hole coherent states but not in the position-momentum phase
space function $h_{\otimes,\rho}^{(\mathrm{red})}(\alpha,\alpha^*)$
built on product coherent states, where the incoming and outgoing
contributions overlap strongly. On the other side the same parameters
also imply an apreciable electron-electron reflection where the
reflected electron has opposite momentum to the incoming. This weak
effect can only be seen in the position-momentum representation (the
weak stripe near $p=-p_F$).

The Andreev reflection from an inhomogeneous superconductor may be
modelled by a pair potential of the form $\Delta(q)=\nu x \theta(x)$.
The stationary solutions at energy $E$ are oscillatory for
$|E|<|\Delta(q)|$ and decay for $|E|>|\Delta(q)|$. We will choose $E\ge
0$ and $\nu>0$. Then the corresponding turning point is at $q_T=\nu/E$.
Figure \ref{fig3} shows reduced phase space distributions for
stationary scattering states for an incoming electronic wave. While
generally the phase space function
$h_{\Join,\rho}^{(\mathrm{red})}(\alpha,\alpha^*)$ in position-velocity
space gives more details than the position-momentum function
$h_{\otimes,\rho}^{(\mathrm{red})}(\alpha,\alpha^*)$. However, the
position-velocity representation does not offer nice semiclassical
descriptions in terms of trajectories. As can be seen in the second row
of Figure~\ref{fig3} these only come about in the position-momentum
representation.

\section{Conclusion}

We have constructed electron-hole coherent states as minimum
uncertainty states for position and velocity for the Bogoliubov-de
Gennes equation. We derived and analysed their main properties: they
entangle space and quasi-spin degrees of freedom and they form an
overcomplete set with an explicit resolution of unity. Basic
applications to stationary scattering in superconductors revealed
both their usefulness and some limitations. In spite of these
limitations we have shown that electron-hole coherent states have a
potential of describing the dynamics of electron-hole excitations in a superconductor
as remain localized in position-velocity phase space for a certain
time. A more detailed description of the dynamics of these states may
lead to new insights on well-known effects in superconductors and
at normalconducting-superconducting interfaces.

\begin{acknowledgments}
  SG would like to thank Martin Zirnbauer for valuable discussions
  in the very early stages of this work.
\end{acknowledgments}


\begin{thebibliography}{99}
\bibitem{schrodinger26} E.~Schr{\"o}dinger.
\newblock Der stetige { {\"U}}bergang von der {M}ikro- zur {M}akromechanik.
\newblock {\em Naturwissenschaften}, 14(28):664--666, 1926.

\bibitem{glauber63} R.~J.~Glauber.
\newblock Coherent and incoherent states of the radiation field.
\newblock {\em Phys. Rev.}, 131(6):2766, 1963.

\bibitem{glauber63a} R.~J.~Glauber.
\newblock The quantum theory of optical coherence.
\newblock {\em Phys. Rev.}, 130(6):2529, 1963.

\bibitem{sudarshan63} E.~C.~G.~Sudarshan.
\newblock Equivalence of semiclassical and quantum mechanical descriptions of
  statistical light beams.
\newblock {\em Phys. Rev. Lett.}, 10(7):277, 1963.

\bibitem{perelomov12} A.~Perelomov.
\newblock {\em Generalized coherent states and their applications}.
\newblock Springer Science \& Business Media, 2012.

\bibitem{klauder85} J.~R.~Klauder and B.-S. Skagerstam, editors.
\newblock {\em Coherent {S}tates. {A}pplications in {P}hysics and
  {M}athematical {P}hysics}.
\newblock World {S}cientific, Singapore, 1985.

\bibitem{ali14} S.~T.~Ali, J.-P.~Antoine, J.-P.~ Gazeau.
\newblock {\em Coherent States, Wavelets, and Their Generalizations.}
2nd edition.
\newblock  Springer, 2014.

\bibitem{gazeau09} J.-P.~ Gazeau.
\newblock {\em Coherent States in Quantum Physics.}
\newblock Wiley VCH, 2009.

\bibitem{combescure12} M.~ Combescure, D.~Robert.
\newblock {\em Coherent States and Applications in Mathematical
  Physics}
\newblock Springer, 2012.

\bibitem{gilmore72} R.~Gilmore.
\newblock Geometry of symmetrized states.
\newblock {\em Ann. Phys.}, 74(2):391 -- 463, 1972.

\bibitem{perelomov72} A.~M.~Perelomov.
\newblock Coherent states for arbitrary lie group.
\newblock {\em Commun. Math. Phys.}, 26(3):222--236, 1972.

\bibitem{radcliffe71} J.~M.~Radcliffe.
\newblock Some properties of coherent spin states.
\newblock {\em J. Phys. A: Gen. Phys.}, 4(3):313, 1971.

\bibitem{arecchi72} F.~T.~Arecchi, E.~Courtens, R.~Gilmore, and
    H.~Thomas.
\newblock Atomic coherent states in quantum optics.
\newblock {\em Phys. Rev. A}, 6(6):2211, 1972.

\bibitem{delbourgo77} R.~Delbourgo.
\newblock Minimal uncertainty states for the rotation and allied groups.
\newblock {\em J. of Phys. A: Math. Gen.}, 10(11):1837,
  1977.

\bibitem{delbourgo77a} R.~Delbourgo and J.~R.~Fox.
\newblock Maximum weight vectors possess minimal uncertainty.
\newblock {\em J. of Phys. A: Math. Gen.}, 10(12):L233,
  1977.

\bibitem{GnLL} S.~Gnutzmann, J.~Langham-Lopez.
\newblock {\em in preparation.}

\bibitem{deGennes66} P.~G.~de~Gennes
\newblock {\em Superconductivity Of Metals And Alloys.}
\newblock Addison-Wesley, 1966.

\bibitem{tinkham04} M.~Tinkham
\newblock {\em Superconductivity}, 2nd rev. edition.
\newblock Dover, 2004.


\end{thebibliography}
\end{document}